\PassOptionsToPackage{unicode}{hyperref}
\PassOptionsToPackage{hyphens}{url}
\PassOptionsToPackage{dvipsnames,svgnames,x11names}{xcolor}
\documentclass[
  12pt]{article}

\usepackage{amsmath,amssymb, amsthm, amsfonts, bm}
\usepackage{iftex}
\ifPDFTeX
  \usepackage[T1]{fontenc}
  \usepackage[utf8]{inputenc}
  \usepackage{textcomp} 
\else 
  \usepackage{unicode-math}
  \defaultfontfeatures{Scale=MatchLowercase}
  \defaultfontfeatures[\rmfamily]{Ligatures=TeX,Scale=1}
\fi
\usepackage{lmodern}
\ifPDFTeX\else  
\fi
\IfFileExists{upquote.sty}{\usepackage{upquote}}{}
\IfFileExists{microtype.sty}{
  \usepackage[]{microtype}
  \UseMicrotypeSet[protrusion]{basicmath} 
}{}
\makeatletter
\@ifundefined{KOMAClassName}{
  \IfFileExists{parskip.sty}{%
    \usepackage{parskip}
  }{
    \setlength{\parindent}{0pt}
    \setlength{\parskip}{6pt plus 2pt minus 1pt}}
}{
  \KOMAoptions{parskip=half}}
\makeatother
\usepackage{xcolor}
\makeatletter
\ifx\paragraph\undefined\else
  \let\oldparagraph\paragraph
  \renewcommand{\paragraph}{
    \@ifstar
      \xxxParagraphStar
      \xxxParagraphNoStar
  }
  \newcommand{\xxxParagraphStar}[1]{\oldparagraph*{#1}\mbox{}}
  \newcommand{\xxxParagraphNoStar}[1]{\oldparagraph{#1}\mbox{}}
\fi
\ifx\subparagraph\undefined\else
  \let\oldsubparagraph\subparagraph
  \renewcommand{\subparagraph}{
    \@ifstar
      \xxxSubParagraphStar
      \xxxSubParagraphNoStar
  }
  \newcommand{\xxxSubParagraphStar}[1]{\oldsubparagraph*{#1}\mbox{}}
  \newcommand{\xxxSubParagraphNoStar}[1]{\oldsubparagraph{#1}\mbox{}}
\fi
\makeatother

\usepackage{longtable,booktabs,array}
\usepackage{calc} 
\usepackage{etoolbox}
\makeatletter
\patchcmd\longtable{\par}{\if@noskipsec\mbox{}\fi\par}{}{}
\makeatother
\IfFileExists{footnotehyper.sty}{\usepackage{footnotehyper}}{\usepackage{footnote}}
\makesavenoteenv{longtable}
\usepackage{graphicx}
\makeatletter
\def\maxwidth{\ifdim\Gin@nat@width>\linewidth\linewidth\else\Gin@nat@width\fi}
\def\maxheight{\ifdim\Gin@nat@height>\textheight\textheight\else\Gin@nat@height\fi}
\makeatother
\setkeys{Gin}{width=\maxwidth,height=\maxheight,keepaspectratio}
\makeatletter
\def\fps@figure{htbp}
\makeatother

\makeatletter
\@ifpackageloaded{caption}{}{\usepackage{caption}}
\AtBeginDocument{%
\ifdefined\contentsname
  \renewcommand*\contentsname{Table of contents}
\else
  \newcommand\contentsname{Table of contents}
\fi
\ifdefined\listfigurename
  \renewcommand*\listfigurename{List of Figures}
\else
  \newcommand\listfigurename{List of Figures}
\fi
\ifdefined\listtablename
  \renewcommand*\listtablename{List of Tables}
\else
  \newcommand\listtablename{List of Tables}
\fi
\ifdefined\figurename
  \renewcommand*\figurename{Figure}
\else
  \newcommand\figurename{Figure}
\fi
\ifdefined\tablename
  \renewcommand*\tablename{Table}
\else
  \newcommand\tablename{Table}
\fi
}
\@ifpackageloaded{float}{}{\usepackage{float}}
\floatstyle{ruled}
\@ifundefined{c@chapter}{\newfloat{codelisting}{h}{lop}}{\newfloat{codelisting}{h}{lop}[chapter]}
\floatname{codelisting}{Listing}

\makeatother
\makeatletter
\@ifpackageloaded{caption}{}{\usepackage{caption}}
\@ifpackageloaded{subcaption}{}{\usepackage{subcaption}}
\makeatother

\ifLuaTeX
  \usepackage{selnolig}  
\fi
\usepackage[]{natbib}
\usepackage{bookmark}

\IfFileExists{xurl.sty}{\usepackage{xurl}}{} 
\hypersetup{
  pdftitle={Regularized Fingerprinting with Linearly Optimal Weight
    Matrix in Detection and Attribution of Climate Change},
  pdfauthor={Haoran Li; Yan Li},
  pdfkeywords={measurement error; linear shrinkage estimator; optimal fingerprinting},
  colorlinks=true,
  linkcolor={blue},
  filecolor={Maroon},
  citecolor={Blue},
  urlcolor={Blue},
  pdfcreator={LaTeX via pandoc}}

\usepackage{afterpage}
\usepackage{appendix}
\usepackage{fullpage}
\usepackage{makecell}
\usepackage{mathtools}
\usepackage[font=small,labelfont=bf,labelsep=space,singlelinecheck=false]{caption}

\newcommand{\anon}{1}

\theoremstyle{plain}
\newtheorem{theorem}{Theorem}
\newtheorem{lemma}{Lemma}
\newtheorem{proposition}[theorem]{Proposition}
\theoremstyle{definition}
\newtheorem{assumption}{Assumption}

\allowdisplaybreaks

\newcommand{\hSigma}{\hat{\bm{\Sigma}}}
\newcommand{\calN}{\mathcal{N}}
\newcommand{\tr}{\mathrm{tr}}
\newcommand{\mE}{\mathbb{E}}

\begin{document}

\def\spacingset#1{\renewcommand{\baselinestretch}%
{#1}\small\normalsize} \spacingset{1}


\if1\anon
{
  \title{\bf Regularized Fingerprinting with Linearly Optimal Weight
    Matrix in Detection and Attribution of Climate Change}  
  \author{{Haoran Li$^1$} and {Yan Li$^1$} \\\\
    $^1$Department of Mathematics and Statistics, Auburn
    University, AL}
  \date{}
  \maketitle
} \fi

\if0\anon
{
  \bigskip
  \bigskip
  \bigskip
  \begin{center}
    {\LARGE\bf Regularized Fingerprinting with Linearly Optimal Weight
  Matrix in Detection and Attribution of Climate Change}
\end{center}
  \medskip
} \fi

\bigskip
\begin{abstract}
  Climate change detection and attribution play a central role in
  establishing the causal influence of human activities on global
  warming. The dominant framework, optimal fingerprinting, is a linear
  errors-in-variables model in which each covariate is subject to
  measurement error with covariance proportional to that of the
  regression error. The reliability of such analyses depends
  critically on accurate inference of the regression coefficients. The
  optimal weight matrix for estimating these coefficients is the
  precision matrix of the regression error, which is typically unknown
  and must be estimated from climate model simulations. However,
  existing regularized optimal fingerprinting approaches often yield
  underestimated uncertainties and overly narrow confidence intervals
  that fail to attain nominal coverage, thereby compromising the
  reliability of analysis. In this paper, we first propose consistent
  variance estimators for the regression coefficients within the class
  of linear shrinkage weight matrices, addressing undercoverage in
  conventional methods. Building on this, we derive a linearly optimal
  weight matrix that directly minimizes the asymptotic variances of
  the estimated scaling factors. Numerical studies confirm improved
  empirical coverage and shorter interval lengths. When applied to
  annual mean temperature data, the proposed method produces narrower,
  more reliable intervals and provides new insights into detection and
  attribution across different regions.
\end{abstract}

\noindent%
\vfill

\newpage
\spacingset{1.8} 


\section{Introduction}
\label{sec:intro}

Successive assessments by the Intergovernmental Panel on Climate
Change (IPCC) have firmly established that more than half of the
observed increase in global average surface temperature in recent
decades can be attributed to anthropogenic increases in greenhouse gas
concentrations and other human-induced forcings
\citep{hegerl2007understanding, Bind:etal:dete:2013,
  eyring2021human}. Detection and attribution (D\&A) analyses have
played a central role in reaching these conclusions.  In climate
science, detection refers to the process of demonstrating that a
climate variable has changed in a statistically significant manner,
without necessarily identifying the cause of that change. Attribution,
on the other hand, involves assessing the extent to which observed
changes can be attributed to multiple external forcings, along with an
assignment of statistical confidence
\citep[e.g.,][]{Hege:Zwie:use:2011}. By comparing climate model
simulations with observed climate variables, detection and attribution
analyses evaluate whether observed changes are statistically
consistent with expected responses, also known as fingerprints or
signals, to one or more external forcings in the climate system.

Optimal fingerprinting (OF), the most widely used method in detection
and attribution analyses, is a multiple linear regression framework in
which observed climate variables are regressed onto the fingerprints
of external forcings \citep[e.g.,][]{hegerl1996detecting,
  Alle:Tett:chec:1999, Alle:Stot:esti:2003}. The primary target of
statistical inference in OF is the vector of regression coefficients,
commonly referred to as scaling factors. These coefficients scale the
fingerprints to match the observed climate changes best. An ideal
point estimator of the scaling factors should be unbiased and exhibit
minimal variance. Confidence intervals constructed around these
estimates quantify uncertainty and form the basis for drawing
detection and attribution conclusions. Specifically, if the confidence
interval for a scaling factor lies significantly above zero, the
effect of the corresponding external forcing is said to be
``detected'' in the observational data. If, in addition, the interval
contains one, then there is sufficient statistical evidence to
``attribute'' the observed changes to that external
forcing. Additionally, a proper confidence interval should have a
coverage rate matching the nominal confidence level to ensure reliable
and statistically robust conclusions.

Historically, OF was deemed ``optimal'' in the context of generalized
least squares (GLS), where the precision matrix of the regression
error is used as a weight for prewhitening. This approach yields
estimators of the scaling factors with minimum variance under
idealized assumptions. However, it was later recognized that
fingerprints are not directly observed but are instead estimated from
climate model simulations, thereby introducing measurement errors and
giving rise to an errors-in-variables (EIV) model. Under the standard
assumption that internal climate variability in model simulations
mirrors that in observations, the errors in the estimated fingerprints
inherit the same covariance structure as the regression errors. As a
result, the estimation framework shifted toward total least squares
(TLS) \citep{Alle:Stot:esti:2003} with both the response and
covariates are ``prewhitened'' using the covariance matrix
\( \Sigma \) of the regression errors.  In practice, \( \Sigma \) is
typically unknown and must be estimated from ensembles of climate
model simulations \citep{Alle:Stot:esti:2003,
  Ribe:Plan:Terr:appl:2013}. This estimation step is generally handled
separately from the regression analysis and treated as a preliminary
step \citep[e.g.,][]{hannart2014optimal}.

Estimating \( \Sigma \) poses significant challenges due to spatial
and temporal dependencies and high dimensionality of the climate
variables. The number of available control runs from climate model
simulations is often insufficient to yield a reliable estimate of
\( \Sigma \). In particular, when the number of control runs is
smaller than the dimension, the sample covariance matrix becomes
singular and cannot be directly inverted to construct a weight matrix.
Early methods addressed this issue by projecting the data onto the
leading empirical orthogonal functions (EOFs) of internal climate
variability, as represented by the empirical covariance matrix of the
control runs \citep{hegerl1996detecting, Alle:Tett:chec:1999}. Later,
\citet{ribes2009adaptation} proposed using a linear shrinkage
estimator of \( \Sigma \) developed by \citet{ledoit2004well}, leading
to the regularized optimal fingerprinting (ROF) method. Confidence
intervals for the scaling factors in ROF can be constructed using
normal approximation techniques \citep{Ribe:Plan:Terr:appl:2013,
  delsole2019, li2021confidence} or bootstrap methods
\citep{delsole2019}. Numerical studies by
\citet{Ribe:Plan:Terr:appl:2013} demonstrated that ROF yields a more
robust and accurate implementation of optimal fingerprinting than the
EOF-based approach, primarily because the performance of EOF methods
is highly sensitive to the selection of the number of retained EOFs.

To address the limitations of EOF truncation,
\citet{katzfuss2017bayesian} introduced a Bayesian framework that
treats the number of retained EOFs as a parameter, allowing robust
inference through averaging over different truncation levels. However,
the Bayesian approach can be computationally intensive, particularly
when the maximum number of truncations is large, and its performance
can be sensitive to the choice of prior distributions. As an
alternative, \citet{hannart2016integrated} proposed an integrated
likelihood formulation that derives a closed-form joint likelihood of
the observational data and control runs by integrating out the unknown
covariance matrix. This integrated approach is equivalent to a
Bayesian model with an informative conjugate prior on the covariance
matrix. It also enables shrinkage toward structured targets beyond the
identity matrix, such as spatio-temporal covariance structures, as
explored in the numerical studies of \citet{hannart2016integrated}. In
practice, however, the prior structure on the covariance matrix may be
uncertain or misspecified, potentially limiting the reliability of the
resulting inference.

The use of an estimated covariance matrix \( \Sigma \) has significant
implications for optimal fingerprinting, especially when the
estimation is based on relatively small samples. Recent studies have
shown that the resulting scaling factor estimator under existing
regularized optimal fingerprinting (ROF) is no longer optimal in terms
of mean squared error (MSE) when \( \Sigma \) is estimated with
substantial uncertainty \citep{li2023regularized}. In such settings,
alternative weight matrices may yield more accurate estimators of the
scaling factors in terms of MSE. Moreover, confidence intervals
constructed using standard practices, such as asymptotic normal
approximations based on two independent samples
\citep{hegerl1996detecting, Alle:Stot:esti:2003} or bootstrap
techniques \citep{delsole2019}, often inadequately account for the
uncertainty in the estimated \( \Sigma \). As a result, the coverage
rates of these intervals are typically lower than the nominal level.
To address this issue, \citet{li2021confidence} proposed a parametric
bootstrap calibration method that inflates the confidence intervals to
achieve nominal coverage. However, this method is computationally
intensive and may perform poorly when the sample size for estimating
\( \Sigma \) is limited.

An alternative approach was developed by \citet{ma2023optimal}, who
proposed an estimation procedure based on estimating equations
(EE). This method produces unbiased estimators that correct for the
bias induced by EIV and constructs weights using the known structure
of the error covariance matrix. The EE approach assumes temporal
stationarity in climate variability, which implies a block Toeplitz
structure for the covariance matrix, and uses a pseudobootstrap
algorithm for constructing confidence intervals. This method improves
both the efficiency of the estimator and the accuracy of the coverage
rates. However, it may lose some efficiency in point estimation due to
not directly estimating the temporal correlation and relying on a
suboptimal weight matrix. In contrast, traditional fingerprinting
methods make use of the full spatio-temporal covariance structure,
which may maximize estimation efficiency but at the cost of large
uncertainty in estimating the high dimensional covariance matrix,
thereby compromising optimality.

To avoid confusion, we henceforth distinguish regularized optimal
fingerprinting (ROF) from regularized fingerprinting (RF) for clarity
in the sequel. This distinction raises a natural and important
question:
\emph{Can optimality in terms of MSE be recovered for scaling factor
  estimators in regularized fingerprinting, at least within a suitably
  defined class of covariance estimators, when the error covariance
  matrix \( \Sigma \) is fully estimated?}

In this paper, we tackle the challenging problems in the regularized
fingerprinting framework based on a class of linear shrinkage
estimators towards an identity matrix of the form
$\hSigma(\lambda) = S + \lambda I$, where $S$ is the sample covariance
matrix from controlled climate simulations and $\lambda>0$ is a tuning
parameter. The contributions of our framework are twofold: First, we
establish the asymptotic properties of the resulting RF estimators of
the scaling factors when the weight matrix is constructed from
\( \hSigma(\lambda) \), under suitable regularity conditions.  We then
propose a data-driven procedure to consistently estimate the
asymptotic covariance matrix of the scaling factor estimators, using
only the estimated fingerprints and the sample covariance matrix.
This directly addresses the undercoverage issue frequently encountered
in regularized fingerprinting, where the uncertainty introduced by
estimating the error covariance matrix is not fully accounted for,
resulting in overly narrow confidence intervals. Second, we introduce
a computational procedure for selecting the optimal regularization
parameter $\lambda$ via grid search, aiming to minimize the total
uncertainty in the scaling factor estimates. The method achieves
asymptotic optimality within the class of linear shrinkage estimators,
as both the sample size and the matrix dimension tend to infinity at a
fixed ratio, potentially less than one. We refer to the proposed
approach as \emph{regularized fingerprinting with linear optimality}.
Extensive simulation studies under realistic conditions demonstrate
that our method provides accurate uncertainty quantification and leads
to confidence intervals with empirical coverage rates close to the
nominal level. Moreover, the optimally selected weight matrix produces
substantially shorter intervals than existing approaches. These
improvements have important practical implications, as precise
uncertainty quantification is central to robust detection and
attribution. In a real-world application to the detection and
attribution of global near-surface air temperature over the period
1951--2020, our method consistently outperforms competing approaches
in terms of interval length and inferential reliability, and provides
new insights into detection and attribution conclusions across
different regional scales. It is worth noting that, although developed
in the context of climate detection and attribution, the proposed
methodology is broadly applicable to multivariate errors-in-variables
regression problems where the covariates are contaminated by
structured noise and the error covariance must be estimated from
limited auxiliary data.

The rest of the paper is organized as follows. In
Section~\ref{sec:methodology}, we briefly review the OF framework and
the asymptotic properties of the class of linear shrinkage estimators
of $\Sigma$, then propose a consistent estimate of the asymptotic
covariance and an optimal weight matrix that minimizes the total
variances of the resulting scaling factors. A large-scale numerical
study assessing the performance of the proposed method is reported in
Section~\ref{sec:numeric}. In Section~\ref{sec:real}, we apply the
proposed method to a detection and attribution analysis of changes in
mean near-surface temperatures on continental and subcontinental
scales. A discussion concludes in Section~\ref{sec:disc}. To improve
readability, we relegate technical details, including proofs of the
theoretical results, additional results from simulation studies, and
details of the climate models, to the Supplementary Materials.

\section{Methodology}\label{sec:methodology}

Fingerprinting can be formulated as a linear regression problem with errors-in-variables (EIV):
\begin{align}
Y &= \sum_{i=1}^p X_i \beta_i + \epsilon, \quad \epsilon \sim \calN(0,\Sigma), \label{eq:fingerprint1} \\
\tilde{X}_{ik} &= X_i + \eta_{ik}, \quad k = 1, \dots, n_i, \label{eq:fingerprint2}
\end{align}
where \( Y \in \mathbb{R}^{N} \) is the observed climate variable of interest, \( X_i \in \mathbb{R}^N \) is the true but unobserved fingerprint of the \( i \)th external forcing, and \( \beta_i \) is the associated scaling factor. The error term \( \epsilon \) is assumed to follow a multivariate normal distribution with mean zero and covariance matrix \( \Sigma \). For each \( i \), \( \tilde{X}_{ik} \) is the $k$th simulation of the fingerprint $X_i$ from a climate model, contaminated by model-specific internal variability \( \eta_{ik} \). The number of ensemble members for the \( i \)th forcing is denoted by \( n_i \). Under the standard assumption that the internal variability in model simulations is consistent with that in observations, the errors \( \eta_{ik} \) are modeled as independent Gaussian noise with covariance \( \Sigma \), i.e., 
\[\eta_{ik} \stackrel{iid}{\sim} \calN(0, \Sigma), ~~j=1,\dots, n_i, ~i=1,\dots, p.\] 
Moreover, $\epsilon$ is independent of $\eta_{ik}$. Both the response \( Y \) and the covariates are assumed to be centered with respect to a common reference period, so no intercept is included in the regression model.

Naturally, the ensemble mean
\[
\tilde{X}_i = \frac{1}{n_i} \sum_{k=1}^{n_i} \tilde{X}_{ik}
\]
serves as an estimator for the true fingerprint \( X_i \). For notational convenience, define
\[
\tilde{X} = (\tilde{X}_1, \dots, \tilde{X}_p), \quad X = (X_1, \dots, X_p), \quad D = \operatorname{diag}(1/n_1, \dots, 1/n_p).
\]
The primary objective is to estimate the scaling factors \( \beta = (\beta_1, \dots, \beta_p)^\top \in \mathbb{R}^p \) and to construct reliable confidence intervals for inference, given the observed response vector \( Y \), the ensemble-averaged fingerprints \( \tilde{X} \), and the scaling matrix \( D \).

\subsection{Optimal Fingerprinting under Idealized Assumptions}\label{subsec:idealized_assumption}

Historically, the term ``optimal'' in optimal fingerprinting originates from early formulations that relied on two idealized assumptions: (i) the error covariance matrix \( \Sigma \) is known, and (ii) the fingerprint matrix \( X \) is fully observed. Under these conditions, the frameworks of \emph{weighted linear regression} and \emph{generalized least squares} (GLS) motivate a prewhitening procedure, in which both \( Y \) and \( X \) are premultiplied by \( \Sigma^{-1/2} \). Then, the GLS estimator of \( \beta \) with weight matrix $\Sigma$ is given by
\[
\hat{\beta}_{\mathrm{GLS}} = (X^T \Sigma^{-1} X)^{-1} X^T \Sigma^{-1} Y.
\]
The optimality of \( \hat{\beta}_{\mathrm{GLS}} \) follows from the fact that it is the best linear unbiased estimator (BLUE) of \( \beta \) under the classical Gauss–Markov assumptions \citep{chen2024statistical}.

Subsequent developments recognized that the true fingerprints \( X \) are not directly observed; instead, only their estimates \( \tilde{X} \) are available. It has been shown that substituting \( \tilde{X} \) for \( X \) in the regression induces bias in the estimation of \( \beta \) \citep{Alle:Stot:esti:2003}. Given knowledge of \( \Sigma \), and under the assumption that the simulation error \( \eta_{ik} \) and the observational error \( \epsilon \) share the same covariance structure, both \( Y \) and \( \tilde{X} \) can be prewhitened accurately.
In this setting, the method of \emph{total least squares} (TLS) can be applied to the prewhitened data. Specifically, the TLS estimator of \( \beta \) with weight matrix $\Sigma$ is given by
\[
\hat{\beta}_{\mathrm{TLS}}(\Sigma) = \arg \min_{\beta} \frac{\|\Sigma^{-1/2} (Y - \tilde{X} \beta)\|_2^2}{1 + \beta^T D \beta},
\]
where \( \|\cdot\|_2 \) denotes the Euclidean norm. For further details, see \citet{gleser1981estimation} and \citet{Alle:Stot:esti:2003}.

Other approaches, such as the estimating equation estimator proposed by \citet{ma2023optimal}, have also been developed, but are beyond the scope of the present work.

\subsection{Regularized Weight Matrix with Linear Shrinkage}\label{subsec:regularized_weight_matrix_with_lienar_shrinkage}

Optimal fingerprinting assumes that the error covariance matrix \( \Sigma \) is known; however, in practice, it must be estimated. The standard approach proceeds in two steps. In the first step, preindustrial control runs from climate model simulations are used to estimate \( \Sigma \). These control runs are generated under the assumption that the models capture only the internal variability of the climate system, without the influence of external forcings. Specifically, suppose we have \( m \) independent, centered control runs \( Z_1, \dots, Z_m \in \mathbb{R}^N \). Under this modeling assumption,
\[
Z_i \stackrel{iid}{\sim} \mathcal{N}(0, \Sigma), \quad i =1,\dots, m.
\]
An estimator of \( \Sigma \) is then constructed from \( Z_1, \dots, Z_m \). In the second step, the estimated covariance matrix is used to prewhiten both the response \( Y \) and the covariates \( \tilde{X} \) in the regression model. The TLS estimator of \( \beta \) is then computed from the prewhitened data.

Nonetheless, consistently estimating \( \Sigma \) in the first step is a challenging task. The traditional estimator is the sample covariance matrix,
\[
S = \frac{1}{m} \sum_{j=1}^m Z_j Z_j^T.
\]
Since \( \Sigma \) is an \( N \times N \) matrix, it has \( N(N+1)/2 \) free parameters when no structural assumptions are imposed. This number is typically too large relative to the available sample size \( m \), which in practice is often at most a few hundred. From a theoretical perspective, in the high-dimensional asymptotic regime where both \( N \) and \( m \) grow and \( N \) is comparable to or exceeds \( m \), the sample covariance matrix \( S \) is known to be inconsistent \citep[]{silverstein1995empirical}.  More severely, when \( N > m \), \( S \) is singular and cannot be directly used as a weight matrix in regression.

As a consequence, the optimality of the original fingerprinting estimator, which relies on the true \( \Sigma \), is unlikely to hold when \( \Sigma \) is replaced by \( S \). In such settings, the statistical properties of the resulting estimator of \( \beta \) are fundamentally governed by the behavior of the estimated weight matrix.

To address the challenge of estimating \( \Sigma \) in high dimensions, \citet{ribes2009adaptation} introduced the linear shrinkage framework developed by \citet{ledoit2004well} into the context of fingerprinting. Specifically, they considered an estimator of $\Sigma$  of the form
\[
\hat{\Sigma}_{\mathrm{LS}} = \rho S + \lambda I_N,
\]
where the scalar tuning parameters \( \rho \) and \( \lambda \) are chosen to minimize the expected mean squared error between the estimator and the true covariance matrix \( \Sigma \). Numerical studies by \citet{ribes2009adaptation} demonstrated that this regularized estimator yields robust and stable results, particularly in settings where the sample size is limited relative to the dimension. The linear shrinkage family can be viewed as an application of classical ridge regularization in the context of fingerprinting. This approach has also been employed in high-dimensional linear hypothesis testing, as in \citet{Li2020high} and \citet{li2020adaptable}.

However, recent findings by \citet{li2023regularized} have shown that the selection of \( \rho \) and \( \lambda \) in the shrinkage estimator is not optimal with respect to MSE between \( \hat{\beta}_{\mathrm{TLS}} \) and the true scaling factor \( \beta \), particularly when the number of control runs \( m \) is relatively small. Moreover, confidence intervals constructed using either normal approximation or bootstrap techniques tend to exhibit low empirical coverage rates, often falling short of the nominal level.

The current work aims to recover the optimality in terms of MSE of the TLS estimator \( \hat{\beta}_{\mathrm{TLS}} \) within the family of linearly regularized covariance estimators. In addition, we aim to develop reliable and computationally efficient procedures for constructing confidence intervals for the scaling factors, based on the optimally regularized estimator.

In particular, our research is formulated as follows. Consider the family of linearly regularized estimators of \( \Sigma \) of the form
\[
\hSigma(\lambda) = S + \lambda I_N,
\]
where \( \lambda > 0 \) is a regularization parameter. The total least squares (TLS) estimator of \( \beta \) based on \( \hat{\Sigma}(\lambda) \) is defined as
\[
\hat{\beta}(\lambda) = \arg \min_{\beta} \frac{\|\hSigma^{-1/2}(\lambda) (Y - \tilde{X} \beta)\|_2^2}{1 + \beta^T D \beta}.
\]
Notably, in contrast to \citet{ribes2009adaptation}, we fix \( \rho = 1 \) in the shrinkage formulation, as the TLS estimator \( \hat{\beta}_{\mathrm{TLS}} \) with weight matrix \( \Sigma(\lambda) \) is invariant to a positive scalar multiplication of \( \hat{\Sigma}(\lambda) \); that is, for any \( a > 0 \), replacing \( \hSigma(\lambda) \) with \( a \hSigma(\lambda) \) yields the same estimate.

Our objective is to find the value of \( \lambda \) that minimizes the asymptotic MSE of \( \hat{\beta}(\lambda) \), i.e.
\[  \lambda_{\mathrm{opt}}  =  \arg\min_{\lambda>0} \lim_{N,m\to\infty} \mE \|\hat{\beta}(\lambda) - \beta\|^2_2,\]
under the high-dimensional asymptotic regime where $N$ and $m$ grow and the ratio $N/m \to c$ for some $c>0$.
In addition, we aim to provide a reliable and computationally efficient estimator of the asymptotic covariance matrix of \( \hat{\beta}(\lambda_{\mathrm{opt}}) \). Based on this estimator, confidence intervals for the scaling factors \( \beta \) can be constructed in a principled and data-driven manner.


\subsection{Optimal Regularization and Confidence Intervals}\label{sec:optimal_shrinkage}

The following assumptions are imposed for the analysis of the asymptotic properties of $\hat{\beta}(\lambda)$.

\begin{assumption}\label{assump:boundedness}
There exist constants $\underline{\alpha}$ and $\overline{\alpha}$ such that $0< \underline{\alpha} < \liminf_{N\to\infty} \ell_{\min}(\Sigma) \leq \limsup_{N\to\infty} \ell_{\max}(\Sigma) < \overline{\alpha}$, where $\ell_{\min}(\cdot)$ and $\ell_{\max}(\cdot)$ are the smallest and largest eigenvalue of a matrix, respectively.
\end{assumption}

\begin{assumption}\label{assump:limit_XX}
Assume that $\lim_{N\to\infty} X^TX/N$ exists and is a positive definite matrix.     
\end{assumption}

\begin{assumption}\label{assump:delta1}
For any fixed $\lambda$, $\lim_{N,m\to \infty} X^T \hSigma^{-1}(\lambda) X/N  = \Delta_1(\lambda)$ exists, where $\Delta_1(\lambda)$ is a nonsingular matrix. 
\end{assumption}

\begin{assumption}\label{assump:delta2}
For any fixed $\lambda$, $\lim_{N,m\to\infty}  X^T \hSigma^{-1}(\lambda) \Sigma \hSigma^{-1}(\lambda) X/N =\Delta_2(\lambda)$ exists, where $\Delta_2(\lambda)$ is a nonsingular matrix.
\end{assumption}

\begin{assumption}\label{assump:2}
For any fixed $\lambda$, $\lim_{N,m\to \infty} \tr{[\hSigma^{-1}(\lambda) \Sigma]}/N$ exists and is a positive constant. 
\end{assumption}

\begin{assumption}\label{assump:K}
For any fixed $\lambda$, $\lim_{N,m\to\infty} \tr{[\hSigma^{-1}(\lambda)\Sigma\hSigma^{-1}(\lambda)\Sigma]}/N =K(\lambda)$ exists with $K(\lambda)>0$. 
\end{assumption}

\begin{theorem}\label{thm:asymptotic_normality}
Under Assumptions \ref{assump:boundedness}--\ref{assump:K}, as $N,m \to \infty$ with $N/m \to c$ for some $c>0$, 
\[\sqrt{N} (\hat\beta(\lambda) - \beta)  \stackrel{\mathcal{D}}{\longrightarrow} \mathcal{N}(0, \Xi(\lambda)),~~\mbox{where}\]
\[ \Xi(\lambda) = (1+\beta^TD\beta) \Delta_1^{-1}(\lambda) \Big\{ \Delta_2(\lambda) + K(\lambda) (D^{-1} + \beta \beta^T)^{-1} \Big\}\Delta_1^{-1}(\lambda).\]
\end{theorem}
Note that the asymptotic MSE of \( \hat{\beta}(\lambda) \) satisfies
\[
N \, \mathbb{E} \| \hat{\beta}(\lambda) - \beta\|_2^2 \longrightarrow \operatorname{tr}(\Xi(\lambda)), \quad \text{as } N, m \to \infty \text{ with } N/m \to c.
\]
The optimal regularization parameter \( \lambda_{\mathrm{opt}} \) is then the minimizer of \( \operatorname{tr}(\Xi(\lambda)) \). We now propose a consistent estimator of $\Xi(\lambda)$ that depends only on the simulated fingerprint matrix \( \tilde{X} \) and the sample covariance matrix \( S \).

Define
\begin{align*}
\Theta_1(\lambda) &= \frac{1 - \lambda Q_1(\lambda)}{1 - (N/m)\{1 - \lambda Q_1(\lambda)\}}, \\
\Theta_2(\lambda) &= \frac{1 - \lambda Q_1(\lambda)}{\left[1 - (N/m)\{1 - \lambda Q_1(\lambda)\}\right]^3} 
- \lambda \frac{Q_1(\lambda) - \lambda Q_2(\lambda)}{\left[1 - (N/m)\{1 - \lambda Q_1(\lambda)\}\right]^4},
\end{align*}
where
\[
Q_1(\lambda) = \frac{1}{N} \operatorname{tr} \left[\hSigma^{-1}(\lambda)\right], \quad 
Q_2(\lambda) = \frac{1}{N} \operatorname{tr} \left[\hSigma^{-2}(\lambda)\right].
\]
Further define
\[
G_1(\lambda) = \frac{1}{N} \tilde{X}^T \hSigma^{-1}(\lambda) \tilde{X}, \quad 
G_2(\lambda) = \frac{1}{N} \tilde{X}^T \hSigma^{-2}(\lambda) \tilde{X}.
\]
Henceforth, $o_{\prec}(1)$ denotes a matrix $A$ such that $\|A\|_F = o_{\mathcal{P}}(1)$, where $\|\cdot \|_F$ denotes the Frobenius norm of a matrix.

\begin{proposition}\label{prop:delta1}
Under Assumptions \ref{assump:boundedness}--\ref{assump:K}, as $N,m \to \infty$ with $N/m \to c$ for some $c>0$, for any fixed $\lambda$,
\[\frac{1}{N} X^T \hSigma^{-1}(\lambda) X = G_1(\lambda) - \Theta_1(\lambda) D + o_{\prec}(1).\]
\end{proposition}

\begin{proposition}\label{prop:delta2}
Under Assumptions \ref{assump:boundedness}--\ref{assump:K}, as $N,m \to \infty$ with $N/m \to c$ for some $c>0$, for any fixed $\lambda$,
\[  \frac{1}{N} X^T \hSigma^{-1}(\lambda) \Sigma \hSigma^{-1}(\lambda) X ={[1+(N/m)\Theta_1(\lambda)]^2} \left[ G_1(\lambda) - \lambda G_2(\lambda) \right] -\Theta_2(\lambda) D + o_{\prec}(1).\]
\end{proposition}
The following result is proved in Lemma 2 of \cite{chen2011regularized}.
\begin{proposition}\label{prop:K1}
Under Assumptions \ref{assump:boundedness}--\ref{assump:K}, as $N,m \to \infty$ with $N/m \to c$ for some $c>0$, for any fixed $\lambda$. 
\[ \tr{[\hSigma^{-1}(\lambda)\Sigma\hSigma^{-1}(\lambda)\Sigma]}/N  = \Theta_2(\lambda) + o_{\mathcal{P}}(1),\]
\end{proposition}

The estimators of the key parameters are then constructed as:  
\begin{align*}
&\hat\Delta_1(\lambda) = G_1(\lambda) - \Theta_1(\lambda)D,\\
&\hat{\Delta}_2(\lambda) =  [1+ (N/m) \Theta_1(\lambda)]^2 \left\{ G_1(\lambda) - \lambda G_2(\lambda)\right\} -\Theta_2(\lambda)D,\\
&\hat{K}(\lambda) = \Theta_2(\lambda),\\
&\hat{\Xi}(\lambda) = \Big(1+\hat{\beta}^T(\lambda)D\hat\beta(\lambda)\Big) \hat\Delta_1^{-1}(\lambda) \Big\{ \hat\Delta_2(\lambda) + \hat{K}(\lambda) \Big(D^{-1} + \hat\beta(\lambda) \hat\beta^T(\lambda)\Big)^{-1} \Big\}\hat\Delta_1^{-1}(\lambda).
\end{align*} 
\begin{lemma}\label{lemma:consistency_estimated_variance}
Under Assumptions \ref{assump:boundedness}--\ref{assump:K}, as $N,m \to \infty$ with $N/m \to c$ for some $c>0$,  for any fixed $\lambda>0$,
\[\hat{\Xi}(\lambda) - \Xi(\lambda) = o_{\prec}(1).\]
\end{lemma}
In practice, the empirical optimal regularization parameter is selected by 
\[
\hat{\lambda}_{\mathrm{opt}} = \arg\min_{\lambda \in [\underline{\lambda}, \overline{\lambda}]} \operatorname{tr}[\hat{\Xi}(\lambda)],
\]
where \( \underline{\lambda} \) and \( \overline{\lambda} \) are prespecified lower and upper search bounds, respectively. Given the eigen-decomposition of \( S \), the computational complexity of evaluating \( \hat{\Delta}_1(\lambda) \), \( \hat{\Delta}_2(\lambda) \), and \( \hat{K}(\lambda) \) for each \( \lambda \) is \( O(pN) \). This allows the optimization over \( \lambda \) to be performed efficiently via a grid search. In finite-sample settings, the quantities \( Q_1(\lambda) \) and \( Q_2(\lambda) \) may become unstable when \( N > n \) and \( \lambda \) is close to zero, due to the near-singularity of \( \hSigma(\lambda) \). To mitigate this issue, it is recommended that the lower bound \( \underline{\lambda} \) of the search range is not too small. A practical choice for the search interval is $[0.01\bar{\tau}, 10\bar{\tau}]$, where \( \bar{\tau} = (1/N) \operatorname{tr}(S) \).

Lastly, we propose marginal confidence intervals for each individual scaling factor \( \beta_i \), as well as a joint confidence region for the vector \( \beta \),  at the asymptotic confidence level \( (1 - \alpha) \).
\begin{itemize}
\item \emph{Marginal confidence interval for \( \beta_i \):}
\[
\beta_i \in \left[ \hat{\beta}_i(\hat\lambda_{opt}) \pm \frac{z_{1 - \alpha/2}}{\sqrt{N}}  {\hat{\Xi}^{1/2}_{ii}(\hat\lambda_{opt})}\right],
\]
where \( z_{1 - \alpha/2} \) denotes the upper \( (1 - \alpha/2) \)-quantile of the standard normal distribution, and \( \hat{\Xi}_{ii}(\hat\lambda_{opt}) \) is the \( i \)th diagonal element of the estimated asymptotic covariance matrix.

\item \emph{Joint confidence region for \( \beta \):}
\[
\left\{ \beta \in \mathbb{R}^p ~:~ \beta^T \hat{\Xi}^{-1}(\hat\lambda_{opt}) \beta \leq N^{-1} \chi^2_p(1 - \alpha) \right\},
\]
where \( \chi^2_p(1 - \alpha) \) is the upper \( (1 - \alpha) \)-quantile of the $\chi^2$-distribution with \( p \) degrees of freedom.
\end{itemize}



\section{Simulation Studies}
\label{sec:numeric}

To evaluate the finite sample performance of the proposed method in comparison with existing practices in Regularized Fingerprinting, we conducted extensive simulation studies emulating realistic settings for detection and attribution analyses of global mean temperature changes, following the
settings of \citet{li2023regularized}. The climate variable of interest consisted of 11 decadal near-surface mean temperatures over 25 spatial grid boxes, resulting in a response vector of dimension $N = 275$.

In each setting, we first set the true fingerprints $X$. Two external forcings were considered, anthropogenic (ANT) and natural (NAT) forcings. The expected fingerprints of these two forcings, denoted by $X_1$ and $X_2$, respectively, were set to the average of all runs from the CNRM-CM5 model simulations, as in \citet{Ribe:Plan:Terr:appl:2013} and adopted in \citet{li2023regularized}. To vary the strength of the signals, each $X_i$ was scaled by $\gamma \in {1, 0.5}$, representing two signal-to-noise regimes. The case of $\gamma=1$ mimics a
global scale study with strong signal strength, while $\gamma=0.5$ matches regional scale studies where the signals are weaker. The true scaling factors were set to $\beta_1= \beta_2 =1$.

For the true covariance matrix $\Sigma$, two spatiotemporal structures were evaluated. The first $\Sigma$ was set to be an unstructured matrix $\Sigma_{UN}$ obtained by manipulating the minimum variance estimate with a set of CNRM-CM5 model simulations considered, as used in \citet{li2023regularized}. The resulting covariance structure resembles the pattern of an unstructured spatial-temporal covariance matrix with variance stationarity and weak dependence over the time dimension considered by \citep{hannart2016integrated}. The second structure, denoted as $\Sigma_{ST}$, was set to be a separable spatiotemporal covariance matrix, where the diagonals were set to be the sample variances from the climate model simulations without imposing temporal stationarity, and the corresponding correlation matrix was set to be the Kronecker product of a spatial correlation matrix and a temporal correlation matrix, both with autoregressive of order 1 and coefficient 0.1.

With the $X_i$, $\beta_i$, $\Sigma$, observed responses $Y$ and noisy fingerprints $(\tilde{X}_1, \tilde{X}_2)$ were generated from models~\eqref{eq:fingerprint1} and~\eqref{eq:fingerprint2}. The regression errors $\epsilon$ followed a multivariate normal distribution $\calN(0, \Sigma)$. The distribution of the measurement error $\eta_i$ for $\tilde X_i$, $i \in \{1,2\}$, was $\calN(0, n_i^{-1}\Sigma)$ with $(n_1, n_2) = (35,46)$, consistent with the number of simulation runs
in a two-way detection and attribution analysis of the annual mean temperature conducted by \citet{Ribe:Plan:Terr:appl:2013}. Control runs $Z_1, \ldots, Z_m$ were generated independently from $\calN(0, \Sigma)$ with sample size $m\in\{50, 100, 200, 400\}$. Here $m =50$ is typical in OF studies, and $m \ge 200$ is possible but not easily obtained unless runs from different climate models are pooled, ignoring the model structure differences.

For each combination of $\gamma$, $\Sigma$ and $m$, we performed 1000 simulation replicates to evaluate the performance of the proposed method with optimally selected tuning parameter $\lambda$, denoted as ``Optim'', in comparison with two existing ROF methods based on TLS. The first competitor, denoted as ``LS-CB'', adopts the ROF method of \citet{Alle:Stot:esti:2003} with a linear shrinkage estimator $\hat \Sigma_{\mathrm{LS}}$ by \citet{ledoit2004well} for prewhitening (LS) and
a calibration bootstrap (CB) for interval estimation \citep{li2021confidence}. The second, denoted as ``MV-CB'', applied the minimum variance estimator $\hat \Sigma_{\mathrm{MV}}$ from \citet{li2023regularized} with the same CB adjustment for constructing confidence intervals. For the proposed method ``Optim'', we constructed the confidence interval from the newly proposed asymptotic results. Since $\hat\Sigma_{\mathrm{LS}}$ falls into the class of linear shrinkage estimators defined in Section~\ref{sec:methodology}, our asymptotic results also apply to this setting. As an additional benchmark, we included an uncalibrated version of the ROF method with $\hat \Sigma_{\mathrm{LS}}$, denoted as ``LS''. The corresponding confidence intervals were constructed from the same normal approximation as our proposed method. Confidence intervals are essential in detection and attribution studies. Ideally, they should be as short as possible while maintaining empirical coverage rates close to the nominal level. To assess performance, we focus on two key metrics, empirical coverage rate and interval length.

\begin{figure}[tbp]
  \centering \includegraphics[width=5.6in]{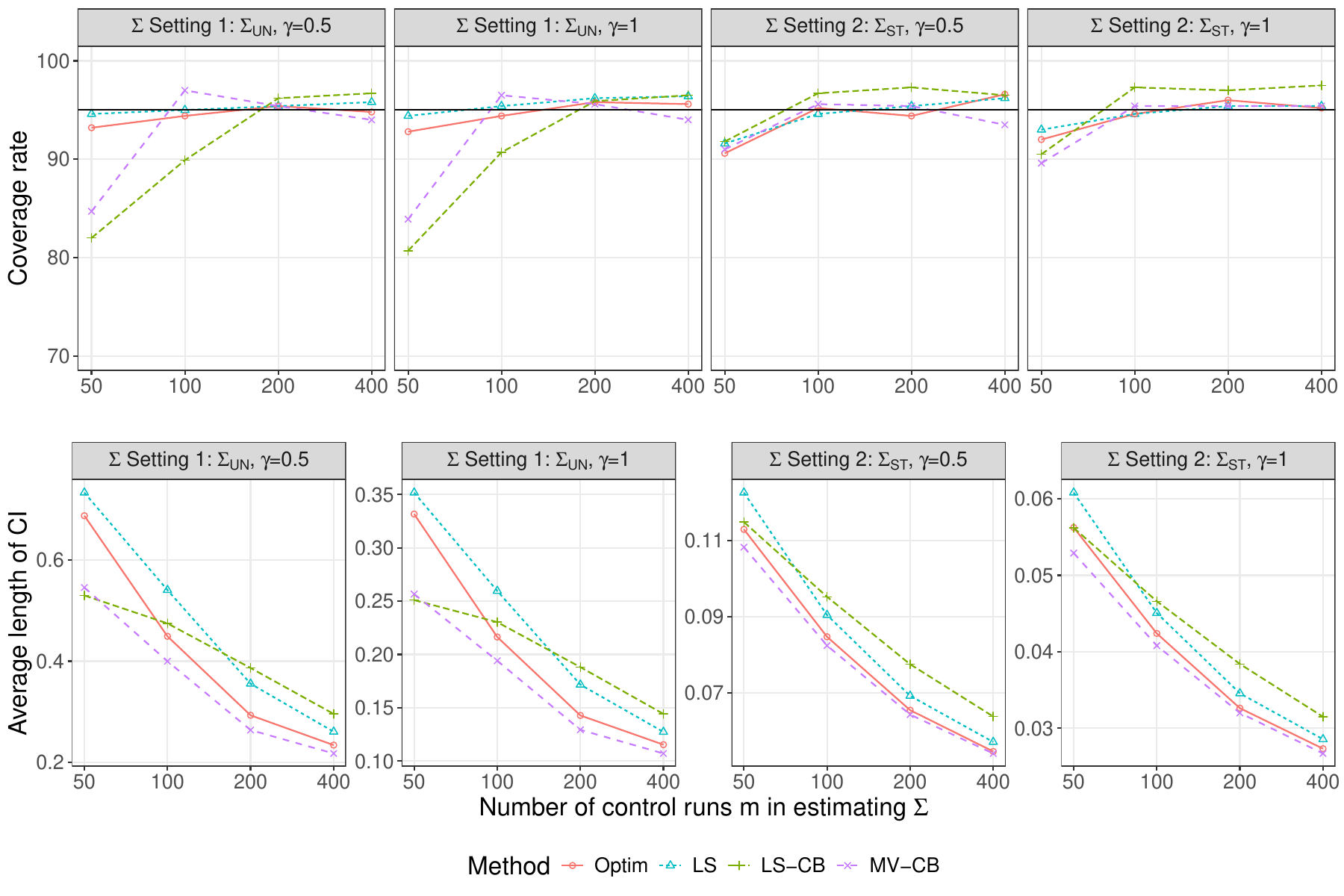}
  \caption{Estimated coverage rates and lengths of 95\% confidence intervals for the ANT scaling factor constructed from four methods, Optim, LS, LS-CB, and MV-CB, based on 1000 replicates. The number of ensembles for estimating the ANT and NAT signals are $n_1 = 35$ and $n_2 = 46$, respectively. The $\gamma$ controls the signal-to-noise ratio for the model. The case of $\gamma=1$ indicates strong signal strength commonly seen in global scale studies, and $\gamma=0.5$ represents a weaker signal case matching with regional scale studies.}
  \label{fig:ci}
\end{figure}

Here, we present in Figure~\ref{fig:ci} the empirical coverage rates
and average lengths of the 95\% confidence intervals derived from the
four competing methods: Optim, LS, LS-CB, and MV-CB, for the ANT
forcing, which is usually the main concern in detection and
attribution analyses of climate changes. Full numerical results for
both ANT and NAT forcings are provided in Table~S1 of the
Supplementary Material, which demonstrates that all methods yield
unbiased point estimates of the scaling factors. For confidence
intervals, our proposed method for estimating the asymptotic variance
of the scaling factors estimators leads to empirical coverage rates
consistently close to the nominal 95\% level across nearly all
settings. Even in the most challenging case with only $m=50$ control
runs, the coverage rate remains around 91\%, supporting the accuracy
of our estimated asymptotic variance. In contrast, both calibrated
competitors (LS-CB and MV-CB) still suffer from undercoverage issues
in low-sample settings such as $m=50$, particularly under an
unstructured covariance matrix $\Sigma_{UN}$. Their coverage improves
as $m$ increases and approaches the nominal level around $m=100$ for
MV-CB and $m = 200$ for LS-CB, an optimistic size in real applications
where the sample becomes comparable to the dimension of
$\Sigma$. Regarding interval length, the Optim method produces
substantially narrower intervals than LS-CB in all comparable
scenarios with desired coverage rate, as is expected due to the
optimal choice of tuning parameter $\lambda$ in the class of linear
shrinkage estimators. Compared to the MV-CB method using a nonlinear
shrinkage estimator, our method remains competitive, particularly in
cases of moderate- to high-sample size where the coverage rate is
close to the nominal level. Notably, MV-CB only provides visibly
shorter intervals than Optim in settings of $\Sigma_{UN}$, which is a
covariance structure favoring the minimum variance estimator. The
interval lengths between Optim and MV-CB are otherwise comparable, as
shown in the lower panel of Figure~\ref{fig:ci}. Overall, our proposed
Optim method offers valid confidence intervals with near-nominal
coverage and competitive or superior interval widths across a wide
range of realistic scenarios. In addition, it also offers substantial
computational advantages over calibration-based approaches, as it
estimates the asymptotic variance without requiring any bootstrap
procedure.

\section{Fingerprinting Mean Temperature Changes}
\label{sec:real}

To demonstrate the performance of proposed methods in real-world
applications, we conducted a detection and attribution analysis of
changes in the mean near-surface air temperature at global (GL),
continental and subcontinent scales over the year period 1951--2020,
utilizing the latest available climate observations and
simulations. Following \citet{zhang2006multimodel} and
\citet{li2023regularized}, we considered several regions: at the
continental scale, Northern Hemisphere (NH), NH midlatitudes (NHM)
between $30^\circ$ and $70^\circ$, Eurasia (EA), and North America
(NA); and at the subcontinental scale, Western North America (WNA),
Central North America (CNA), and Eastern North America (ENA), where
spatio-temporal correlation structures are more likely to
hold. Detection and attribution analyses for two external forcings,
anthropogenic (ANT) and natural (NAT) forcings, were conducted for
each region.

For each regional analysis,
Models~\eqref{eq:fingerprint1}--\eqref{eq:fingerprint2} require three
components: the observed mean temperature $Y \in \mathbb{R}^N$, the
estimated fingerprints $\tilde{X}_{ANT}$ and $\tilde{X}_{NAT}$, and
independent control runs $Z_1, \ldots, Z_m$ for estimating the
covariance matrix $\Sigma$.

\subsection{Data Preparation}

We first obtained the observational vector $Y$ from the latest
HadCRUT5 dataset \citep{morice2021updated}, which provides monthly
anomalies of near-surface air temperature from January 1850 on
$5^\circ \times 5^\circ$ grid boxes relative to the 1961--1990
reference period. At each grid box, annual anomalies were computed
from monthly values provided that at least nine months of data were
available within a given year; otherwise, the annual mean was marked
as missing. Nonoverlapping 5-year averages were subsequently
calculated, requiring no more than two missing annual values within
each 5-year period. After removing the 1961--1965 period due to
centering, 13 values of 5-year averages were obtained per grid box.

To reduce spatial dimensionality for the global and continental-scale
analyses, available $5^\circ \times 5^\circ$ grid boxes were
aggregated into larger spatial resolutions. In particular, grid box
sizes were set to $40^\circ \times 30^\circ$ for GL and NH,
$40^\circ \times 10^\circ$ for NHM, $10^\circ \times 20^\circ$ for EA,
and $10^\circ \times 5^\circ$ for NA. For subcontinental regional
analyses on WNA, CNA and ENA, the original $5^\circ \times 5^\circ$
grid boxes were maintained to preserve finer spatial detail. A summary
of the spatiotemporal dimensions, specifically the number of grid
boxes, number of time steps, and the total count of observations after
handling missing values, is provided in Table~\ref{tab:regions}.

\begin{table}[tbp]
  \footnotesize
  \centering
  \caption{\sf Summaries of the names, coordinate ranges, ideal
    spatio-temporal dimensions ($S$ and $T$), and dimension of
    observation after removing missing values of the 5 regions
    analyzed in the study.}
  \label{tab:regions}
  \def\arraystretch{1}\tabcolsep=0.4em
  \begin{tabular}{llcccccc}
    \toprule
    Acronym & Regions & Longitude  & Latitude   & Grid size & $S$ & $T$ & $N$ \\
            &         & ($^\circ$E) & ($^\circ$N) & ($1^\circ \times 1^\circ$) & & &\\
    \midrule
    \multicolumn{8}{c}{Global and Continental Regions}\\
    GL & Global & $-$180 / 180 & $-$90 / 90 & $40 \times 30$ & 54 & 13 & 696 \\
    NH & Northern Hemisphere & $-$180 / 180 & 0 / 90 & $40 \times 30$ & 27 & 13 & 352 \\
    NHM & Northern Hemisphere $30^\circ N$ to $70^\circ N$ & $-$180 / 180 & 30 / 70 & $40 \times 10$ & 36 & 13 & 468 \\
    EA & Eurasia & $-$10 / 180 & 30 / 70 & $10 \times 20$ & 38 & 13 & 494 \\
    NA & North America & $-$130 / $-$50 & 30 / 60 & $10 \times 5$ & 48 & 13 & 624 \\
    \addlinespace[1ex]
    \multicolumn{8}{c}{Subcontinental Regions}\\
    \midrule
    WNA & Western North America & $-$130 / $-$105 & 30 / 60 & $5 \times 5$ & 30 & 13 & 390 \\
    CNA & Central North America & $-$105 / $-$85 & 30 / 50 & $5 \times 5$ & 16 & 13 & 208 \\
    ENA & Eastern North America & $-$85 / $-$50 & 15 / 30 & $5 \times 5$ & 21 & 13 & 273 \\
    \bottomrule
  \end{tabular}
\end{table}

We obtained estimated fingerprints and control runs using outputs from
CMIP6 multimodel simulations \citep{eyring2016overview} for the
selected period of 1951--2020. These simulations included hist-GHG
experiments (driven exclusively by changes in well-mixed greenhouse
gas concentrations), hist-aer experiments (driven exclusively by
changes in anthropogenic aerosol emissions and burdens), hist-nat
experiments (driven exclusively by natural forcings), as well as
preindustrial control simulations of varying durations representing
internal climate variability. Details of the climate model simulations
are summarized in Table~S2 of the Supplementary Material.

In particular, the fingerprints $\tilde{X}_{NAT}$ for the NAT forcing
were obtained directly by averaging over 40 available runs. For the
ANT forcing, which is typically the primary focus in climate detection
and attribution studies, direct simulation outputs were not available
from CMIP6 models. Under the linear additivity assumption,
$X_{ANT} = X_{GHG} + X_{AER}$ \citep{zhang2006multimodel}, we
constructed $\tilde{X}_{ANT}$ by combining the greenhouse gas and
aerosol fingerprints for each model. During processing, the same
missing data pattern observed in $Y$ was imposed on the fingerprints,
and the same averaging and spatial aggregation procedures were applied
to reduce dimensionality. Since both the observational and model data
were centered relative to the 1961--1990 period, the first 5-year
block (1961--1965) was excluded to maintain consistency.

The control runs were similarly constructed from 29 preindustrial
control simulations produced by global climate models participating in
the CMIP6 ensemble. The available control simulations varied in length
from approximately 100 to 1200 years. To mitigate the effects of model
drift, a long-term linear trend was removed separately at each grid
box for each control run. To increase the effective sample size for
estimating the covariance matrix $\Sigma$, we assumed temporal
stationarity, a standard practice in climate studies, and split each
control simulation into nonoverlapping 70-year blocks corresponding to
the 1951--2020 analysis period. This yielded a total of $m = 181$
independent replicates for covariance estimation. Each 70-year block
was then subject to the same missing data masking and dimension
reduction procedures applied to $Y$. Details of the number of
available simulations from each preindustrial control dataset are
summarized in Table~S3 of the Supplementary Material.

\subsection{Results}

\begin{figure}[tbp]
  \centering \includegraphics[width=5.6in]{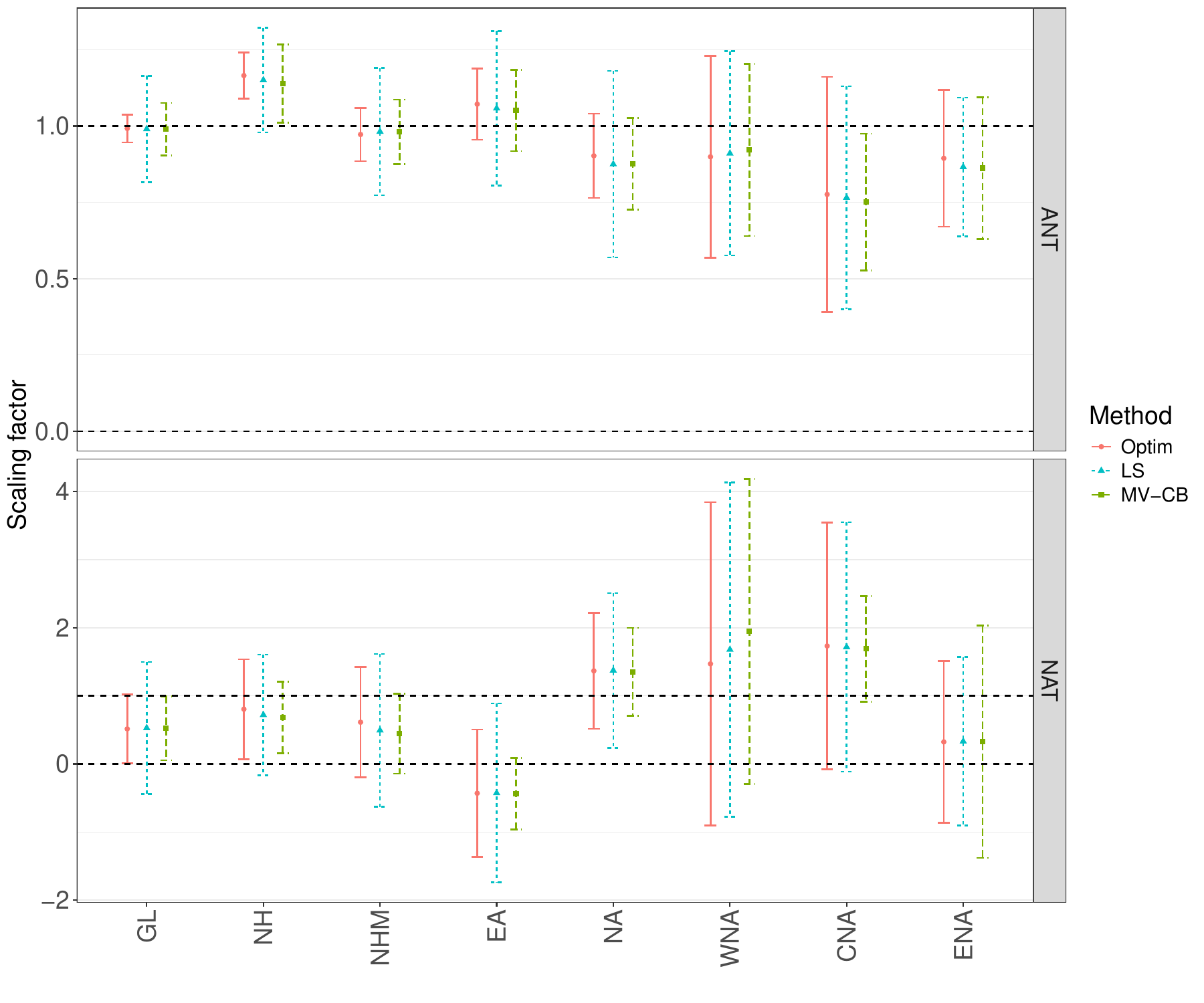}
  \caption{Estimated signal scaling factors for ANT and NAT required
    to best match observed 1950--2020 annual mean temperature for
    different spatial domains, and the corresponding 95\% confidence
    intervals from different methods. For weight matrix construction,
    ``Optim'' denotes the proposed linearly optimal method, ``LS''
    denotes the linear shrinkage estimator and ``MV-CB'' denotes the
    minimum variance estimator with parametric calibration for
    confidence intervals.}
  \label{fig:real_data}
\end{figure}

Figure~\ref{fig:real_data} summarizes the estimated scaling factors
for the two external forcings, ANT and NAT, along with their
associated 95\% confidence intervals, obtained using the proposed
method (``Optim''), the LS method based on the new asymptotic results
(``LS''), and the traditional TLS method using the minimum variance
estimator $\hat \Sigma_{MV}$ (``MV-CB''). Across all analyses, the
point estimates produced by the three methods are generally similar,
particularly for larger-scale regions, though discrepancies appear at
the subcontinental scale such as WNA. Given the demonstrated
robustness of the proposed method in simulation studies, its estimates
are considered more reliable. Notably, the scaling factors for the ANT
forcing are close to one in most regional analyses, suggesting that
the multimodel average of the ANT fingerprints captures the magnitude
of the observed temperature trends over the past 70 years. An
exception is the NH analysis, where the proposed method indicates that
climate model simulations tend to overestimate the expected response
to anthropogenic forcing. Regarding uncertainty quantification in
confidence interval, the proposed method consistently yields narrower
confidence intervals compared to the LS method, and provides intervals
of comparable length (narrower when signal is strong) to those
produced by the MV-CB approach, consistent with findings from the
simulation studies. When the sample size is limited to the dimension
of $\Sigma$, the results of the two existing approaches may be
questionable due to the undercoverage issues previously discussed.

For the detection and attribution conclusions, results are largely
consistent across the three methods for the ANT forcing, with
detection achieved in all regional analyses, albeit with slight
differences in magnitude. In contrast, results for the NAT forcing
show greater variability across methods, owing to the weaker signal
associated with natural influences on climate change
\citep{gillett2021constraining}. In particular, for the CNA region,
the Optim and LS methods lead to detection and attribution conclusions
for the ANT forcing, while the MV-CB method supports detection
only. For the NAT forcing, the MV-CB method suggests detection and
attribution, whereas the Optim and LS methods do not, as their
confidence intervals cover zero. In supercontinent scale analyses,
such as GL and NH, MV-CB yields weaker detection statements compared
to the other two methods, as the confidence intervals are close to
zero. In other regions, the three methods produce consistent
conclusions. Although it is not possible to definitively determine
which method is more accurate for this single analysis, given the
unknown true underlying scaling factors, we emphasize that the
comparative performance of these methods has been thoroughly evaluated
in the simulation studies. Overall, the proposed Optim method
generally provides more reliable results, characterized by lower
uncertainty and higher computational efficiency.

\section{Discussion}
\label{sec:disc}

Optimal fingerprinting, the principal methodological framework for
detection and attribution studies in climate change research, has
substantial influence on contemporary climate analysis. These analyses
provide foundational support for observationally constrained climate
projections and facilitate the estimation of critical climate system
parameters, such as climate sensitivity. Nonetheless, the original
optimality condition of optimal fingerprinting, which involves
minimizing total uncertainty in scaling factor estimators, is
compromised when estimated fingerprints of external forcings are
subject to measurement errors and when the covariance matrix $\Sigma$
of regression errors needs to be estimated rather than known as a
priori. Furthermore, existing TLS approaches used to estimate scaling
factors frequently exhibit undercoverage issues, primarily due to the
underestimation of variances of the resulting scaling factor
estimators.

Our proposed methodology addresses these limitations and offers
improvements in two key aspects. First, within the regularized
fingerprinting framework utilizing a linear shrinkage weight matrix,
we propose an efficient, data-driven procedure for consistently
estimating the asymptotic covariance matrix of the scaling factor
estimators. This procedure utilizes only the estimated fingerprints
and the sample covariance matrix from control runs, enabling the
effective construction of confidence intervals for scaling factors
based on normal approximations. Within the class of linear shrinkage
estimators, our approach yields valid confidence intervals with
close-to-nominal coverage rates at substantially reduced computational
cost compared to the existing calibration method
\citep{li2021confidence}. Unlike methods dependent on explicit
distributional assumptions \citep{hannart2016integrated,
  katzfuss2017bayesian} or those requiring temporal stationarity to
mitigate undercoverage at the expense of efficiency
\citep{ma2023optimal}, our method imposes no additional assumptions
and fully leverages the spatio-temporal covariance structure of
$\Sigma$, thus potentially achieving optimality. Our second
contribution involves determining the optimal weight matrix within the
linear shrinkage estimator class. By directly minimizing the
asymptotic mean squared error (MSE) of the scaling factor estimator,
we identify the optimal shrinkage parameter and construct the
corresponding weight matrix by inverting the resulting optimal linear
shrinkage estimator. This method is demonstrably more efficient than
current regularized fingerprinting practices, as evidenced by both
simulation studies and real-world applications. Consequently, the
original optimality of fingerprinting is substantially restored in
practical terms, reducing uncertainty in key quantities such as
attributable warming and climate sensitivity.

Overall, the proposed method represents a promising, easy-to-implement
detection and attribution tool for practical applications, offering
point estimates with lower MSE and confidence intervals with desirable
coverage rates. By directly addressing the long-standing undercoverage
issue, our methodology improves the reliability of inference. As
demonstrated in our application, while qualitative conclusions
regarding detection and attribution may remain largely consistent with
those from existing methods, revisiting the main results that support
attribution statements in IPCC Assessment Reports
\citep{eyring2021human} using our framework is both feasible and
worthwhile. A publicly available software implementation further
facilitates this task.

The proposed method can be extended in several directions. It is
particularly interesting to investigate how asymptotic results under
conditions where $N,m \to \infty$ and $N/m \to c$ might inform and
refine regularized fingerprinting practices. Adjusting temporal and
spatial resolution, which determines $N$, could significantly affect
the efficiency of inference and thus alter the conclusions of
detection and attribution analyses. Additionally, our current approach
overlooks differences among climate models by treating the runs under
each forcing as identical. In practice, these runs typically originate
from different climate models. A more realistic modeling approach
should explicitly account for the heterogeneity in variability among
different models in estimating fingerprints under each external
forcing. Furthermore, developing a goodness-of-fit procedure to verify
the consistency of variability between observations and climate models
would enhance the reliability of detection and attribution
analyses. Finally, the proposed method is also applicable beyond
climate science. In particular, multivariate errors-in-variables
regression problems with high-dimensional structure and limited
auxiliary data are common in fields such as genomics, econometrics,
and neuroimaging. In these domains, where covariates are measured with
structured noise and precise inference is essential, our regularized
framework offers a principled approach to improving estimator
efficiency and uncertainty quantification. These broader applications
represent a natural direction for future methodological development
and practical deployment.

\bigskip

\section*{Supplementary Materials}

\setcounter{figure}{0}
\setcounter{table}{0}
\setcounter{lemma}{0}

\renewcommand{\theequation}{\Alph{section}.\arabic{equation}}
\renewcommand{\thelemma}{S\arabic{lemma}}
\renewcommand\thefigure{S\unskip\arabic{figure}}
\renewcommand\thetable{S\unskip\arabic{table}}

\begin{appendices}
\section{Proofs of the Main Text}
\label{Sec:proof}

\subsection{Technical Lemmas}
We collect several technical lemmas in this section. In what follows, \( \|\cdot\|_2 \) denotes the operator norm (spectral norm) of a matrix.

\begin{lemma}[Woodbury Matrix Identity]\label{lemma:woodbury}
The following identity holds:
\[
(A + UCV)^{-1} = A^{-1} - A^{-1} U \left( C^{-1} + V A^{-1} U \right)^{-1} V A^{-1},
\]
for matrices \( A, U, C, V \) of conformable sizes, assuming all inverses exist and are well-defined.
\end{lemma}

\begin{lemma}\label{lemma:concentration}
Suppose that $W \sim \calN(0, I_N)$, and let $C$ be a symmetric $N\times N$ matrix with $\|C\|_2 \leq L$. Then, for all $0<t<L$,
\[\mathbb{P} (\frac{1}{N}|W^T CW -\tr(C )| >t ) \leq 2 \exp\left(-\frac{Nt^2}{4L^2} \right).\]
\end{lemma}
The lemma is known in the literature. See, for example, \cite{paul2007asymptotics}.

\begin{lemma}[Lemma 2.7 of \cite{bai1998no}]\label{lemma:burkholder}
Let $W = (w_1, \dots, w_p)^T$, where $w_i$'s  are i.i.d. real r.v.'s with mean $0$ and variance $1$. Let $B$ be a deterministic matrix. Then, for any $m\geq 2$, we have 
\[\mE |W^T BW - \tr (B)|^m \leq C_m (\mE w_1^4 \tr(BB^T))^{m/2} + C_m \mE w_1^{2m} \tr[ (BB^T)^{m/2}],\]
where $C_m$ is a constant only depending on $m$. 
\end{lemma}
We now consider a modified version of \( \hSigma \) in which the dependence on a specific \( Z_k \) is removed, defined as
\[\hSigma_k = \frac{1}{m}\sum_{i\neq k} Z_iZ_i^T + \lambda I_N.\] 
The following result can be shown following similar arguments as those in Lemma 2.10 of \cite{bai1998no}.
\begin{lemma}\label{lemma:residual_Sigma_k_Sigma}
For any matrix $D$ and $\lambda>0$,
\[ |\tr(\hSigma^{-1} C) - \tr (\hSigma^{-1}_k C)| \leq \frac{\| C\Sigma\|_2}{\lambda}.\]
\end{lemma}

\subsection{Preliminary Results} 

We summarize several well-known results from Random Matrix Theory (RMT); for detailed derivations and background, see \citet{bai2010spectral}. Let \( \ell_1, \dots, \ell_N \) denote the eigenvalues of the population covariance matrix \( \Sigma \), and define its empirical spectral distribution as
\[
F^{\Sigma}(\tau) = \frac{1}{N} \sum_{j=1}^N \mathbf{1}(\tau \geq \ell_j).
\]
For any \( z \in \mathbb{C}^+ \coloneqq \{ u + iv \in \mathbb{C} : v > 0 \} \), there exists a unique solution \( \underline{s}(z) \in \mathbb{C}^+ \) to the equation
\[
\underline{s}(z) = \int \frac{dF^{\Sigma}(\tau)}{\tau (1 - c - c z \underline{s}(z)) - z},
\]
commonly referred to as the Mar\v{c}enko–Pastur equation. Here, \( c = \lim_{N \to \infty} N/m \in (0, \infty) \) denotes the limiting aspect ratio.

The function \( \underline{s}(z) \) admits a smooth extension to the negative real axis. In particular, for any \( \lambda \in \mathbb{R}^+ \), the limit
\[
\lim_{v \to 0^+} \underline{s}(-\lambda + iv) = s(-\lambda)
\]
exists and defines a real-valued function \( s(\cdot) \) that is analytic on \( \mathbb{R}^- \). Moreover, \( s(-\lambda) \) satisfies the transformed Mar\v{c}enko–Pastur equation:
\begin{equation}\label{eq:MP_lambda1}
s(-\lambda) = \int \frac{dF^{\Sigma}(\tau)}{\tau (1 - c + \lambda c s(-\lambda)) + \lambda}.
\end{equation}


Recall that $Q_1(\lambda) = N^{-1}\tr(\hSigma^{-1}(\lambda))$ and $Q_2(\lambda) = N^{-1} \tr(\hSigma^{-2}(\lambda))$. It is clear that $Q_2(\lambda) = -Q_1'(\lambda)$. Also recall the definition of $\Theta_1(\lambda)$ and $\Theta_2(\lambda)$ as 
\begin{align*}
&\Theta_1(\lambda) = \frac{1 - \lambda Q_1(\lambda) }{1- (N/m) \{ 1- \lambda Q_1(\lambda) \} },\\
&\Theta_2(\lambda) = \frac{1-\lambda Q_1(\lambda)}{[1 - (N/m)\{1 -\lambda Q_1(\lambda) \}]^3  }   -\lambda \frac{ Q_1(\lambda) -\lambda Q_2(\lambda)}{ [1- (N/m)\{1- \lambda Q_1(\lambda) \}]^4 }\\[5pt]
& \phantom{\Theta_2(\lambda)} = (1+(N/m)\Theta_1(\lambda))^2 \left(\Theta_1(\lambda) + \lambda \Theta_1'(\lambda) \right)  .
\end{align*}

Under Assumptions \ref{assump:boundedness}--\ref{assump:K}, if \( N, m \to \infty \) with \( N/m \to c > 0 \), then for any fixed \( \lambda > 0 \), the quantities \( Q_1(\lambda) \) and \( Q_2(\lambda) \) satisfy
\[
Q_1(\lambda) \stackrel{P}{\longrightarrow} s(-\lambda), \quad \text{and} \quad
Q_2(\lambda) \stackrel{P}{\longrightarrow} s'(-\lambda),
\]
where \( s(-\lambda) \) is the limiting Stieltjes transform defined in \eqref{eq:MP_lambda1}, and \( s'(-\lambda) \) denotes its derivative.

Define
\begin{align*}
\Omega_1(\lambda) &= \frac{1 - \lambda s(-\lambda)}{1 - c \left\{ 1 - \lambda s(-\lambda) \right\}}, \\
\Omega_2(\lambda) &= \frac{1 - \lambda s(-\lambda)}{\left[1 - c \left\{ 1 - \lambda s(-\lambda) \right\} \right]^3}
- \lambda \frac{ s(-\lambda) - \lambda s'(-\lambda)}{\left[1 - c \left\{ 1 - \lambda s(-\lambda) \right\} \right]^4} \\
&= \left(1 + c \Omega_1(\lambda)\right)^2 \left\{ \Omega_1(\lambda) + \lambda \Omega_1'(\lambda) \right\}.
\end{align*}
Here, it is known that \( 1 - c \left\{ 1 - \lambda s(-\lambda) \right\} \) is bounded away from zero for all \( \lambda \in \mathbb{R}^+ \).

Then, it follows directly that
\begin{equation}\label{eq:converge_theta1_theta2}
\Theta_1(\lambda) - \Omega_1(\lambda) \stackrel{P}{\longrightarrow} 0 \quad \text{and} \quad
\Theta_2(\lambda) - \Omega_2(\lambda) \stackrel{P}{\longrightarrow} 0.    
\end{equation}

\subsection{Concentration of Traces of Various Matrices and Quadratic Forms}

The following results are established in Lemma 2 of \citet{chen2011regularized}.

\begin{lemma}\label{lemma:theta1_theta2}
Under Assumptions \ref{assump:boundedness}--\ref{assump:K}, if \( N, m \to \infty \) with \( N/m \to c > 0 \), then for any \( \lambda > 0 \),
\begin{align*}
&\frac{1}{N} \tr\left[\hSigma^{-1}(\lambda) \Sigma\right] - \Omega_1(\lambda) = o_p(1), \\[5pt]
&\frac{1}{N} \tr\left[\hSigma^{-1}(\lambda) \Sigma \hat{\Sigma}^{-1}(\lambda) \Sigma\right] - \Omega_2(\lambda) = o_p(1).
\end{align*}
\end{lemma}
\noindent Note that the second result implies Proposition~\ref{prop:K1} of the manuscript, since \( \Omega_2(\lambda) = \Theta_2(\lambda) + o_p(1) \).

Following the strategy of Lemma 2 in \citet{chen2011regularized}, we can also establish the following analogous result.
\begin{lemma}\label{lemma:theta3}
Under Assumptions \ref{assump:boundedness}--\ref{assump:K}, if \( N, m \to \infty \) with \( N/m \to c > 0 \), then for any \( \lambda > 0 \),
\[
\frac{1}{N} \tr\left[\hSigma^{-2}(\lambda) \Sigma\right] + \Omega_1'(\lambda) = o_p(1).
\]
\end{lemma}
In the following, we outline the key steps in the proof of Lemma~\ref{lemma:theta3}, omitting detailed calculations as they closely follow the arguments used in the proof of Lemma~2 in \citet{chen2011regularized}. Consider the identity 
\[  S+ \lambda I_N - \lambda I_N = \frac{1}{m} \sum_{k=1}^m Z_k Z_k^T.\]
Multiplying both sides by $\hSigma^{-2}(\lambda)$ and taking the averaged trace, we have 
\[ Q_1(\lambda) - \lambda Q_2(\lambda) = \frac{1}{mN}\sum_{k=1}^m Z_k^T \hSigma^{-2}(\lambda)Z_k.\]
Using the Woodbury matrix identity, 
\begin{align*}
     Q_1(\lambda) - \lambda Q_2(\lambda) &= \frac{1}{mN} \sum_{k=1}^m \frac{Z_k^T \hSigma_k^{-2}(\lambda) Z_k}{ (1+ m^{-1}Z_k^T \hSigma^{-1}_k(\lambda)Z_k  )^2 }\\
     & = \frac{1}{mN} \sum_{k=1}^m \frac{Z_k^T\hSigma^{-2}_k(\lambda)Z_k}{( 1+ m^{-1}\tr(\hSigma^{-1}_k(\lambda)\Sigma))^2} + R^{(1)}\\
     & = \frac{1}{m} \sum_{k=1}^m \frac{ N^{-1} \tr(\hSigma^{-2}_k(\lambda)\Sigma)}{ (1+ m^{-1}\tr(\hSigma^{-1}_k(\lambda)\Sigma))^2} + R^{(2)} + R^{(1)}\\
     & = \frac{1}{m}\sum_{k=1}^m \frac{ N^{-1} \tr(\hSigma^{-2}(\lambda)\Sigma)}{ (1+ m^{-1}\tr(\hSigma^{-1}(\lambda)\Sigma))^2} + R^{(3)} + R^{(2)} + R^{(1)}\\
     & = \frac{ N^{-1} \tr(\hSigma^{-2}(\lambda)\Sigma)}{ (1+ m^{-1}\tr(\hSigma^{-1}(\lambda)\Sigma))^2} +R^{(4)} + R^{(3)} + R^{(2)} + R^{(1)}
\end{align*}
for appropriate residual terms $R^{(j)}$, $j=1,2,3,4$. Following the strategy of the proof of Lemma~2 in \citet{chen2011regularized}, we can use Lemma \ref{lemma:concentration}, Lemma \ref{lemma:burkholder}, Lemma \ref{lemma:residual_Sigma_k_Sigma}  to show that
\[
\sup_{1 \leq j \leq 4} R^{(j)} = o_p(1).
\]
The details are omitted, as they closely mirror those in the referenced proof.

It follows then
\[  \frac{1}{N} \tr(\hSigma^{-2}(\lambda)\Sigma) = [1 + (N/m)\Theta_1(\lambda)]^2 (Q_1(\lambda) - \lambda Q_2(\lambda)) + o_p(1) = -\Omega_1'(\lambda) + o_p(1). \]
It completes the proof of Lemma \ref{lemma:theta3}.

\vspace{10pt}
Next, we establish a series of analogous results concerning the concentration of quadratic forms involving \( \hSigma(\lambda) \) and \( \Sigma \). Throughout the remainder of this section, let \( a \in \mathbb{R}^N \) denote a sequence of deterministic vectors satisfying the normalization condition \( N^{-1} \|a\|_2^2 = 1 \). The following three lemmas are shown in \cite{karoui2011geometric}. 
\begin{lemma}
    \label{lemma:deterministic_equivalent1}
    Under Assumptions \ref{assump:boundedness}-\ref{assump:K}, if $N,m\to\infty$ such that $N/m\to c >0$, 
    \[ \frac{1}{N} a^T \hSigma^{-1}(\lambda) a - \frac{1}{N} a^T\left( \frac{1}{1+c\Omega_1(\lambda)} \Sigma + \lambda I_N\right)^{-1}a \stackrel{P}{\longrightarrow} 0.\]
\end{lemma}

\begin{lemma}
    \label{lemma:deterministic_equivalent2}
    Under Assumptions \ref{assump:boundedness}-\ref{assump:K}, if $N,m\to\infty$ such that $N/m\to c >0$, 
    \begin{align*}
    \frac{1}{N}  a^T &\hSigma^{-2}(\lambda) a + \frac{1}{N} a^T \frac{d}{d\lambda}\left( \frac{1}{1+c\Omega_1(\lambda)} \Sigma + \lambda I_N\right)^{-1} a \stackrel{P}{\longrightarrow} 0.
\end{align*}
\end{lemma}

\begin{lemma}
    \label{lemma:deterministic_equivalent3}
    Under Assumptions \ref{assump:boundedness}-\ref{assump:K}, if $N,m\to\infty$ such that $N/m\to c >0$, 
    \begin{align*}
    \frac{1}{N}  a^T &\hSigma^{-1}(\lambda) \Sigma \hSigma^{-1}(\lambda) a \\
    &- \frac{1}{N} \left(1 +\frac{c\Omega_2(\lambda)}{(1+c\Omega_1(\lambda))^2 }\right) a^T\left( \frac{1}{1+c\Omega_1(\lambda)} \Sigma + \lambda I_N\right)^{-1} \Sigma\left( \frac{1}{1+c\Omega_1(\lambda)} \Sigma + \lambda I_N\right)^{-1} a \stackrel{P}{\longrightarrow} 0.
\end{align*}
\end{lemma}

Combining Lemma~\ref{lemma:deterministic_equivalent1}, Lemma~\ref{lemma:deterministic_equivalent2}, and Lemma~\ref{lemma:deterministic_equivalent3}, and using the relationship between \( \Omega_1(\lambda) \) and \( \Omega_2(\lambda) \), we obtain
\begin{equation}\label{eq:concentration_hSigma_Sigma_hSigma}
\frac{1}{N} X^T \hSigma^{-1}(\lambda) \Sigma \hSigma^{-1}(\lambda) X 
- \left(1 + c\, \Omega_1(\lambda)\right)^2 \left( \frac{1}{N} X^T \hSigma^{-1}(\lambda) X 
- \lambda \frac{1}{N} X^T \hSigma^{-2}(\lambda) X \right) 
\stackrel{P}{\longrightarrow} 0.
\end{equation}

\subsection{Proof of Theorem \ref{thm:asymptotic_normality}}\label{proof_theorem_normality}

The theorem is obtained by applying Theorem 2 of \citet{li2023regularized}. Reparameterize the model as
\[
Y = \tilde{X}^* \beta^* + \epsilon,
\]
where \( \tilde{X}^* = \tilde{X} D^{-1/2} \) and \( \beta^* = D^{1/2} \beta \). It is straightforward to verify that all columns of \( \tilde{X}^* \) have the same covariance matrix \( \Sigma \). Let \( \hat{\beta}^*(\lambda) \) denote the TLS estimator of \( \beta^* \) using the weight matrix \( \hat{\Sigma}(\lambda) \). Then, it holds that
\[
\hat{\beta}^*(\lambda) = D^{1/2} \hat{\beta}(\lambda).
\]
The conditions of Theorem 2 in \citet{li2023regularized} are satisfied under this reparameterized model. Applying the theorem yields
\[
\sqrt{N} \left( \hat{\beta}^*(\lambda) - \beta^* \right) \stackrel{\mathcal{D}}{\longrightarrow} \mathcal{N}\left( 0, \Xi^*(\lambda) \right),
\]
where
\begin{align*}
\Xi^*(\lambda) = 
(D^{-1/2} \Delta_1 D^{-1/2})^{-1} 
&\left\{ D^{-1/2} \Delta_2 D^{-1/2} + K \left(I_p + \beta^* (\beta^*)^T\right)^{-1} \right\}\times \\
&\left(1 + (\beta^*)^T \beta^*\right) 
(D^{-1/2} \Delta_1 D^{-1/2})^{-1}.
\end{align*}
The proof of Theorem \ref{thm:asymptotic_normality} is complete by noting that $ D^{-1/2}\Xi^*(\lambda) D^{-1/2} = \Xi(\lambda)$.

\subsection{Proof of Proposition~\ref{prop:delta1} and Proposition~\ref{prop:delta2}}
Propositions~\ref{prop:delta1} and \ref{prop:delta2} are proved by analyzing the bias introduced in the quadratic forms
\[
\frac{1}{N} X^T \hSigma^{-1}(\lambda) X \quad \text{and} \quad \frac{1}{N} X^T \hSigma^{-1}(\lambda) \Sigma \hSigma^{-1}(\lambda) X
\]
when \( X \) is replaced by \( \tilde{X} \).

Let \( W = (W_1, \dots, W_p) \) be an \( N \times p \) matrix with independent and identically distributed \( \mathcal{N}(0,1) \) entries, and assume that \( W \) is independent of \( \epsilon \). Then, the ensemble-averaged fingerprint matrix \( \tilde{X} \) has the same distribution as
\[
\tilde{X} = X + \Sigma^{1/2} W D^{1/2}.
\]
In the following analysis, we shall assume without loss of generality that \( \tilde{X} \) takes the form \( X + \Sigma^{1/2} W D^{1/2} \).

First, consider $G_1(\lambda)= \tilde{X}^T \hSigma^{-1}(\lambda) \tilde{X}/N$.  We decompose  it as 
\begin{align*}
G_1(\lambda)=& \frac{1}{N} X^T \hSigma^{-1}(\lambda) X + \frac{1}{N} D^{1/2} W^T \Sigma^{1/2} \hSigma^{-1}(\lambda) \Sigma^{1/2} W D^{1/2} + \\
&\frac{1}{N} D^{1/2} W^T \Sigma^{1/2} \hSigma^{-1}(\lambda) X + \frac{1}{N} X^T \hSigma^{-1}(\lambda) \Sigma^{1/2} WD^{1/2}.
\end{align*}
For $i=1,2,\dots, p$, using Lemma \ref{lemma:concentration}, Lemma \ref{lemma:theta1_theta2} and Equation \eqref{eq:converge_theta1_theta2},
\[\frac{1}{N}W_i^T  \Sigma^{1/2} \hSigma^{-1}(\lambda) \Sigma^{1/2} W_i  = \frac{1}{N}\tr[\hSigma^{-1}(\lambda)\Sigma] + o_p(1) =\Omega_1(\lambda) +o_p(1) =  \Theta_1(\lambda)  + o_p(1). \]
For $1\leq i\neq j\leq p$, since $W_i$ and $W_j$ are independent, 
\begin{align*}
&\frac{1}{N}\mE W_i^T  \Sigma^{1/2} \hSigma^{-1}(\lambda) \Sigma^{1/2} W_j = 0, \\
&\frac{1}{N^2} \mE (W_i^T \Sigma^{1/2} \hSigma^{-1}(\lambda) \Sigma^{1/2} W_j)^2 = \frac{1}{N^2} \mE \tr(\hSigma^{-1}(\lambda) \Sigma) = O_p(N^{-1}).
\end{align*}
Combining the results, we get
\[ \frac{1}{N}  W^T \Sigma^{1/2} \hSigma^{-1}(\lambda) \Sigma^{1/2} W  = \Theta_1(\lambda) I_p + o_p(1).\]
Similarly, for any $i =1,2,\dots, p$,
\begin{align*} 
&\frac{1}{N}\mE W_i^T \Sigma^{1/2}\hSigma^{-1}(\lambda) X = 0,\\
&\frac{1}{N^2} W_i^T \Sigma^{1/2}\hSigma^{-1}(\lambda) XX^T  \hSigma^{-1}(\lambda)\Sigma^{1/2} W_i = \frac{1}{N^2}\tr[X^T \hSigma^{-1}(\lambda) \Sigma \hSigma^{-1}(\lambda) X ] + o_p(N^{-1})  = O_p(N^{-1}).
\end{align*}
All together, we obtain 
\[G_1(\lambda) = \frac{1}{N}\tilde{X}^T \hSigma^{-1}(\lambda) \tilde{X} = \frac{1}{N} X^T \hSigma^{-1}(\lambda) X  + \Theta_1(\lambda) D + o_{\prec}(1). \]
It complete the proof of Proposition \ref{prop:delta1}.

\vspace{10pt}

Second, consider $G_2(\lambda) = \tilde{X}^T \hSigma^{-2}(\lambda)\tilde{X}/N$. We can decompose it as 
\begin{align*}
G_2(\lambda) =& \frac{1}{N} X^T \hSigma^{-2}(\lambda) X + \frac{1}{N} D^{1/2} W^T \Sigma^{1/2} \hSigma^{-2}(\lambda) \Sigma^{1/2} W D^{1/2} + \\
&\frac{1}{N} D^{1/2} W^T \Sigma^{1/2} \hSigma^{-2}(\lambda) X + \frac{1}{N} X^T \hSigma^{-2}(\lambda) \Sigma^{1/2} WD^{1/2}.
\end{align*}

Following analogous arguments as those in the analysis of $G_1(\lambda)$, for $i=1,2,\dots, p$, using Lemma \ref{lemma:concentration}, Lemma \ref{lemma:theta1_theta2} and Equation \eqref{eq:converge_theta1_theta2},
\[\frac{1}{N}W_i^T  \Sigma^{1/2} \hSigma^{-2}(\lambda) \Sigma^{1/2} W_i  = \frac{1}{N}\tr[\hSigma^{-2}(\lambda)\Sigma] + o_p(1) =- \Omega_1'(\lambda) +o_p(1) = - \Theta_1'(\lambda)  + o_p(1). \]
For $1\leq i\neq j\leq p$, since $W_i$ and $W_j$ are independent, 
\begin{align*}
&\frac{1}{N}\mE W_i^T  \Sigma^{1/2} \hSigma^{-2}(\lambda) \Sigma^{1/2} W_j = 0, \\
&\frac{1}{N^2} \mE (W_i^T \Sigma^{1/2} \hSigma^{-2}(\lambda) \Sigma^{1/2} W_j)^2 = \frac{1}{N^2} \mE \tr(\hSigma^{-2}(\lambda) \Sigma) = O_p(N^{-1}).
\end{align*}
Combining the results, we get
\[ \frac{1}{N}  W^T \Sigma^{1/2} \hSigma^{-2}(\lambda) \Sigma^{1/2} W  = -\Theta_1'(\lambda) I_p + o_{\prec}(1).\]
Similarly, for any $i =1,2,\dots, p$,
\begin{align*} 
&\frac{1}{N}\mE W_i^T \Sigma^{1/2}\hSigma^{-2}(\lambda) X = 0,\\
&\frac{1}{N^2} W_i^T \Sigma^{1/2}\hSigma^{-2}(\lambda) XX^T  \hSigma^{-2}(\lambda)\Sigma^{1/2} W_i = \frac{1}{N^2}\tr[X^T \hSigma^{-2}(\lambda) \Sigma \hSigma^{-2}(\lambda) X ] + o_p(N^{-1})  = O_p(N^{-1}).
\end{align*}
All together, we obtain 
\[G_2(\lambda) = \frac{1}{N}\tilde{X}^T \hSigma^{-2}(\lambda) \tilde{X} = \frac{1}{N} X^T \hSigma^{-2}(\lambda) X  - \Theta_1'(\lambda) D + o_{\prec}(1). \]

\vspace{10pt}

Combine the decomposition of $G_1(\lambda)$ and $G_2(\lambda)$ and Equation \eqref{eq:concentration_hSigma_Sigma_hSigma}.
\begin{align*}
&\frac{1}{N} X^T \hSigma^{-1}(\lambda) \Sigma \hSigma^{-1}(\lambda) X \\
&=  \left(1 + c\, \Omega_1(\lambda)\right)^2 \left( \frac{1}{N} X^T \hSigma^{-1}(\lambda) X 
- \lambda \frac{1}{N} X^T \hSigma^{-2}(\lambda) X \right) +o_{\prec}(1)\\
&= (1+ (N/m) \Theta_1(\lambda))^2 \left(G_1(\lambda) - \Theta_1(\lambda) D- \lambda G_2(\lambda) - \lambda \Theta_1'(\lambda)D \right) + o_{\prec}(1)\\
&= (1+ (N/m) \Theta_1(\lambda))^2 (G_1(\lambda) - \lambda G_2(\lambda)) -  \Theta_2(\lambda)D + o_{\prec}(1).
\end{align*}
It completes the proof of Proposition \ref{prop:delta2}.

\section{Detailed Results on Simulation Studies}
\label{Sec:results}

The results of simulation studies in Section~\ref{sec:numeric} of the
main text are detailed in Table~\ref{Tabap:existing_n_35_46}.
\afterpage{
  {\small
    \spacingset{0.9}
\begin{longtable}{llrrrrrrrr}
  \caption{Scaling factor estimates and 95\% confidence interval. Two
    structures of covariance matrix is considered:
    $\Sigma_{\mathrm{UN}}$ for proposed shrinkage estimator from
    ensemble simulations and $\Sigma_{\mathrm{ST}}$ for separable
    spatio-temporal covariance matrix. $\gamma$ is the scale to
    control the signal-to-noise ratio. Error terms are generated from
    multivariate normal distribution. Number of ensembles for two
    forcings are $n_1=35$ and $n_2=46$. Four methods are compared,
    Optim, LS, LS-CB, and MV-CB, based on 1000 replicates. Bias and
    standard deviation (SD $\times$ 100) of scaling factors and
    average length of confidence intervals (CIL) and corresponding
    empirical rate (CR) from 1000 replicates are
    recorded. } \label{Tabap:existing_n_35_46}\\
  \toprule
  & &  \multicolumn{4}{c}{ANT} & \multicolumn{4}{c}{NAT}\\
  \cmidrule(lr){3-6} \cmidrule(lr){7-10}
  size & method & Bias & SD & CIL & CR & Bias & SD & CIL & CR \\
  \midrule \multicolumn{10}{c}{$\Sigma$ Setting 1:
    $\Sigma_{\mathrm{UN}}$;
    SNR $\gamma = 0.5$}\\
  50 & Optim & -0.00 & 17.8 & 0.69 & 93.2 & -0.00 & 51.2 & 2.00 & 92.0 \\
  & LS & -0.00 & 18.0 & 0.73 & 94.6 & -0.01 & 54.1 & 2.24 & 92.6 \\
  & LS-CB & 0.02 & 19.0 & 0.53 & 82.0 & 0.01 & 57.0 & 1.89 & 89.9 \\
  & MV-CB & 0.02 & 17.3 & 0.55 & 84.7 & 0.01 & 49.7 & 1.79 & 92.6 \\
  100 & Optim & -0.00 & 11.0 & 0.45 & 94.4 & -0.01 & 30.7 & 1.27 & 95.2 \\
  & LS & -0.00 & 13.2 & 0.54 & 95.0 & -0.01 & 36.9 & 1.53 & 95.0 \\
  & LS-CB & 0.01 & 13.6 & 0.47 & 89.9 & -0.02 & 36.9 & 1.46 & 94.0 \\
  & MV-CB & -0.00 & 9.7 & 0.40 & 97.0 & -0.02 & 26.8 & 1.13 & 95.6 \\
  200 & Optim & -0.00 & 6.9 & 0.29 & 95.4 & 0.00 & 20.3 & 0.85 & 95.6 \\
  & LS & -0.00 & 8.3 & 0.36 & 95.4 & 0.00 & 23.3 & 1.01 & 96.0 \\
  & LS-CB & 0.00 & 9.0 & 0.39 & 96.2 & -0.01 & 24.4 & 1.08 & 97.5 \\
  & MV-CB & -0.00 & 6.7 & 0.26 & 95.4 & -0.01 & 18.7 & 0.76 & 94.8 \\
  400 & Optim & -0.00 & 5.4 & 0.23 & 94.8 & 0.00 & 15.6 & 0.68 & 96.0 \\
  & LS & -0.00 & 6.1 & 0.26 & 95.8 & 0.00 & 17.5 & 0.76 & 95.4 \\
  & LS-CB & 0.00 & 6.1 & 0.30 & 96.7 & -0.01 & 18.9 & 0.83 & 97.0 \\
  & MV-CB & -0.00 & 5.6 & 0.22 & 94.0 & -0.00 & 16.1 & 0.64 & 94.3 \\
  \midrule \multicolumn{10}{c}{$\Sigma$ Setting 1:
    $\Sigma_{\mathrm{UN}}$;
    SNR $\gamma = 1$}\\
  50 & Optim & 0.00 & 9.1 & 0.33 & 92.8 & -0.01 & 18.3 & 0.73 & 93.2 \\
  & LS & 0.00 & 9.1 & 0.35 & 94.4 & -0.01 & 19.2 & 0.79 & 94.8 \\
  & LS-CB & 0.00 & 9.0 & 0.25 & 80.7 & -0.01 & 20.4 & 0.73 & 92.4 \\
  & MV-CB & 0.00 & 8.5 & 0.26 & 83.9 & -0.00 & 19.2 & 0.68 & 89.6 \\
  100 & Optim & -0.00 & 5.2 & 0.22 & 94.4 & -0.01 & 12.4 & 0.52 & 94.6 \\
  & LS & -0.00 & 6.1 & 0.26 & 95.4 & -0.01 & 13.6 & 0.58 & 96.2 \\
  & LS-CB & 0.00 & 6.6 & 0.23 & 90.7 & -0.01 & 14.8 & 0.59 & 96.2 \\
  & MV-CB & 0.00 & 4.5 & 0.19 & 96.5 & -0.01 & 12.0 & 0.46 & 94.6 \\
  200 & Optim & 0.00 & 3.3 & 0.14 & 95.8 & -0.01 & 9.5 & 0.38 & 94.0 \\
  & LS & 0.00 & 4.1 & 0.17 & 96.2 & -0.01 & 10.5 & 0.43 & 95.4 \\
  & LS-CB & 0.00 & 4.5 & 0.19 & 95.9 & -0.00 & 10.2 & 0.46 & 97.5 \\
  & MV-CB & 0.00 & 3.3 & 0.13 & 95.6 & 0.00 & 8.7 & 0.35 & 95.9 \\
  400 & Optim & 0.00 & 2.8 & 0.12 & 95.6 & -0.00 & 7.8 & 0.32 & 95.2 \\
  & LS & 0.00 & 3.1 & 0.13 & 96.4 & -0.00 & 8.5 & 0.35 & 95.2 \\
  & LS-CB & 0.00 & 3.2 & 0.14 & 96.5 & -0.00 & 8.3 & 0.37 & 96.7 \\
  & MV-CB & 0.00 & 2.8 & 0.11 & 94.0 & -0.00 & 7.8 & 0.30 & 95.4 \\
  \midrule \pagebreak
    \multicolumn{10}{c}{Table~S1 continued}\\
    \midrule
  & &  \multicolumn{4}{c}{ANT} & \multicolumn{4}{c}{NAT}\\
  \cmidrule(lr){3-6} \cmidrule(lr){7-10}
  size & method & Bias & SD & CIL & CR & Bias & SD & CIL & CR \\
  \midrule
  \multicolumn{10}{c}{$\Sigma$ Setting 2:
    $\Sigma_{\mathrm{ST}}$;
    SNR $\gamma = 0.5$}\\
  50 & Optim & -0.00 & 3.2 & 0.11 & 90.6 & 0.00 & 8.5 & 0.33 & 92.6 \\
  & LS & -0.00 & 3.4 & 0.12 & 91.6 & -0.00 & 8.6 & 0.35 & 94.2 \\
  & LS-CB & -0.00 & 3.2 & 0.11 & 91.8 & 0.00 & 8.7 & 0.39 & 96.7 \\
  & MV-CB & 0.00 & 3.1 & 0.11 & 91.0 & 0.00 & 8.5 & 0.35 & 95.1 \\
  100 & Optim & -0.00 & 2.1 & 0.08 & 95.2 & -0.00 & 6.3 & 0.27 & 95.6 \\
  & LS & -0.00 & 2.2 & 0.09 & 94.6 & -0.00 & 6.5 & 0.28 & 96.4 \\
  & LS-CB & -0.00 & 2.2 & 0.10 & 96.7 & 0.00 & 7.1 & 0.30 & 95.6 \\
  & MV-CB & -0.00 & 2.0 & 0.08 & 95.6 & 0.00 & 7.0 & 0.26 & 91.6 \\
  200 & Optim & -0.00 & 1.6 & 0.07 & 94.4 & -0.00 & 5.1 & 0.22 & 94.0 \\
  & LS & -0.00 & 1.7 & 0.07 & 95.4 & -0.00 & 5.2 & 0.23 & 95.4 \\
  & LS-CB & -0.00 & 1.6 & 0.08 & 97.3 & 0.00 & 5.9 & 0.25 & 96.5 \\
  & MV-CB & -0.00 & 1.6 & 0.06 & 95.4 & 0.00 & 5.6 & 0.21 & 93.7 \\
  400 & Optim & -0.00 & 1.3 & 0.05 & 96.6 & -0.00 & 4.4 & 0.18 & 94.2 \\
  & LS & -0.00 & 1.4 & 0.06 & 96.2 & -0.00 & 4.5 & 0.19 & 95.8 \\
  & LS-CB & -0.00 & 1.5 & 0.06 & 96.5 & 0.00 & 5.1 & 0.21 & 93.7 \\
  & MV-CB & -0.00 & 1.4 & 0.05 & 93.5 & -0.00 & 4.8 & 0.18 & 90.7 \\
  \midrule \multicolumn{10}{c}{$\Sigma$ Setting 2:
    $\Sigma_{\mathrm{ST}}$;
    SNR $\gamma = 1$}\\
  50 & Optim & -0.00 & 1.4 & 0.06 & 92.0 & 0.00 & 4.2 & 0.16 & 91.6 \\
  & LS & 0.00 & 1.5 & 0.06 & 93.0 & 0.00 & 4.3 & 0.16 & 93.0 \\
  & LS-CB & -0.00 & 1.6 & 0.06 & 90.5 & 0.00 & 4.4 & 0.18 & 94.8 \\
  & MV-CB & -0.00 & 1.6 & 0.05 & 89.6 & 0.00 & 4.5 & 0.16 & 91.6 \\
  100 & Optim & 0.00 & 1.1 & 0.04 & 94.6 & 0.00 & 3.5 & 0.13 & 94.4 \\
  & LS & 0.00 & 1.1 & 0.05 & 94.6 & 0.00 & 3.6 & 0.14 & 95.0 \\
  & LS-CB & -0.00 & 1.1 & 0.05 & 97.3 & 0.00 & 3.2 & 0.14 & 97.8 \\
  & MV-CB & -0.00 & 1.0 & 0.04 & 95.4 & 0.00 & 3.2 & 0.12 & 95.9 \\
  200 & Optim & 0.00 & 0.8 & 0.03 & 96.0 & 0.00 & 2.7 & 0.11 & 95.4 \\
  & LS & 0.00 & 0.8 & 0.03 & 95.4 & 0.00 & 2.8 & 0.11 & 95.0 \\
  & LS-CB & -0.00 & 0.8 & 0.04 & 97.0 & 0.00 & 2.8 & 0.12 & 96.7 \\
  & MV-CB & 0.00 & 0.7 & 0.03 & 95.4 & 0.00 & 2.7 & 0.10 & 94.6 \\
  400 & Optim & -0.00 & 0.7 & 0.03 & 95.2 & 0.00 & 2.3 & 0.09 & 96.8 \\
  & LS & -0.00 & 0.7 & 0.03 & 95.4 & 0.00 & 2.4 & 0.10 & 95.2 \\
  & LS-CB & 0.00 & 0.7 & 0.03 & 97.5 & 0.00 & 2.3 & 0.10 & 97.8 \\
  & MV-CB & 0.00 & 0.6 & 0.03 & 95.4 & 0.00 & 2.3 & 0.09 & 94.6 \\
  \bottomrule
\end{longtable}
}}

\section{Details of the CMIP6 climate models}
\label{Sec:cmip6}

The CMIP6 climate model simulations are detailed in Table~\ref{Tab:model} and Table~\ref{Tab:control}.

\afterpage{
  \spacingset{1}
\begin{table}[tbp]
  \centering
  \caption{\sf List of CMIP6 ensembles of simulations used to estimate
    the fingerprint of external forcings. Total numbers for the
    forcings are 43, 43, and 40 for GHG, AER, and NAT
    respectively.} \label{Tab:model}
  \begin{tabular}{ llr | llr}
    \toprule
    Model & Ensemble & \# of Runs & Model & Ensemble & \# of Runs \\
    \midrule
    BCC-CSM2-MR & hist-GHG &  3 &  IPSL   & hist-GHG & 10 \\
          & hist-aer &  3 &         & hist-aer & 10 \\
          & hist-nat &  3 &         & hist-nat & 10 \\
    CanESM5     & hist-GHG & 10 &  MIROC6 & hist-GHG & 3 \\
          & hist-aer & 10 &         & hist-aer & 3 \\
          & hist-nat & 10 & MRI-ESM2& hist-GHG & 3 \\
    CNRM        & hist-GHG & 10 &         & hist-aer & 3 \\
          & hist-aer & 10 &         & hist-nat & 3 \\
          & hist-nat & 10 & NorESM2 & hist-aer & 3 \\
    HadGEM3     & hist-GHG &  4 &         &  &   \\
          & hist-nat &  4 &         &  &   \\
    \bottomrule
  \end{tabular}
\end{table}

\begin{table}[tbp]
\centering
\caption{List of CMIP6 control simulations used to estimate the 
internal climate variability. Number of replicates indicate the number 
of non-overlapping blocks of 70-year segments taken from 
the model simulations} \label{Tab:control}
\begin{tabular}{lrlr}
  \toprule
  Model & \makecell[r]{Number of\\replicates} & Model & \makecell[r]{Number of\\replicates} \\ 
  \midrule
  AWI-CM-1-1-MR & 7 & GISS-E2-1-G (r1f3) & 1 \\ 
  BCC-CSM2-MR & 8 & GISS-E2-1-G (r2f1) & 1 \\ 
  BCC-ESM1 & 6 & GISS-E2-1-G-CC & 2 \\ 
  CAMS-CSM1-0 & 7 & GISS-E2-1-H & 11 \\ 
  CNRM-CM6-1 & 7 & HadGEM3-GC31-LL & 7 \\ 
  CNRM-ESM2-1 & 7 & HadGEM3-GC31-MM & 7 \\ 
  EC-Earth3 & 7 & IPSL-CM6A-LR (r1i1) & 17 \\ 
  EC-Earth3-Veg & 7 & IPSL-CM6A-LR (r1i2) & 3 \\ 
  FIO-ESM-2-0 & 5 & NorCPM1 (r1) & 7 \\ 
  GFDL-CM4 & 7 & NorCPM1 (r2) & 7 \\ 
  GFDL-ESM4 & 7 & NorCPM1 (r3) & 7 \\ 
  GISS-E2-1-G (r101) & 2 & NorESM1-F & 2 \\ 
  GISS-E2-1-G (r102) & 2 & NorESM2-LM & 4 \\ 
  GISS-E2-1-G (r1f1) & 12 & UKESM1-0-LL & 10 \\ 
  GISS-E2-1-G (r1f2) & 4 &  & Total: 181 \\ 
  \bottomrule
\end{tabular}
\end{table}
}

\end{appendices}

\clearpage


\bibliography{cited}

\end{document}